\documentclass[11pt]{article}
\usepackage[english]{babel}
\usepackage{amsmath,amssymb,amsbsy,amstext,amsthm,booktabs,simplewick,exscale,relsize,slashed,graphicx,amsfonts,upgreek,color}
\usepackage{multirow,dcolumn,bm,enumerate,url,hyperref, cite}

\newcommand{\tev}{{\rm TeV}}
\newcommand{\kev}{{\rm keV}}
\newcommand{\gev}{{\rm GeV}}
\newcommand{\mev}{{\rm MeV}}
\newcommand{\dd}{{\rm d}}
\newcommand{\kms}{\rm km/s}

\newcommand{\tlab}{\theta_{\rm lab}}
\usepackage[usenames,dvipsnames]{xcolor}

\newcommand{\be}{\begin{equation}}
\newcommand{\ee}{\end{equation}}

\newcommand{\ltap}{\lesssim}
\newcommand{\eg}{{\it e.g.}}
\newcommand{\ie}{{\it i.e.}}

\begin{document}

\title{\bf The Inelastic Frontier: \\ Discovering Dark Matter at 
High Recoil Energy}

\author{Joseph Bramante$^1$, Patrick J. Fox$^2$, Graham D. Kribs$^3$, 
Adam Martin$^1$ \\[2mm]
$^1$Department of Physics, University of Notre Dame, \\ 
225 Nieuwland Hall, Notre Dame, IN, 46556 USA \\
$^2$Theoretical Physics Department, Fermilab, Batavia, IL, 60510 USA \\
$^3$Department of Physics, University of Oregon, Eugene, OR, 97403 USA}

\date{}
\maketitle
\vspace{-7.5cm}
\noindent \makebox[11.5cm][l]{\small \hspace*{-.2cm} }{\small Fermilab-Pub-16-301-T}  \\  [-1mm]
\vspace{7.5cm}
\begin{abstract}

There exist well motivated models of particle dark matter which predominantly 
scatter inelastically off nuclei in direct detection experiments.
This inelastic transition causes the dark matter to up-scatter
in terrestrial experiments into an excited state up to $550$ keV heavier than the dark matter itself.  
An inelastic transition of this size is highly
suppressed by both kinematics and nuclear form factors. 
In this paper, we extend previous studies of inelastic dark matter
to determine the present bounds on the scattering cross section,
and the prospects for improvements in sensitivity.  
Three scenarios provide illustrative examples: 
nearly pure Higgsino supersymmetric dark matter; 
magnetic inelastic dark matter;
and inelastic models with dark photon exchange.  
We determine the elastic scattering rate 
(through loop diagrams involving the heavy state)
as well as verify that exothermic transitions are negligible
(in the parameter space we consider). 
Presently, the strongest bounds on the
cross section are from xenon at LUX-PandaX 
(when the mass splitting $\delta \lesssim 160$~keV), 
iodine at PICO (when $160 \lesssim \delta \lesssim 300$~keV), 
and tungsten at CRESST (when $\delta \gtrsim 300$~keV).  
Amusingly, once $\delta \gtrsim 200$~keV, weak scale (and larger) 
dark matter - nucleon
scattering cross sections are allowed. 
The relative competitiveness of these diverse experiments
is governed by the \emph{upper} bound on the 
recoil energies employed by each experiment, as well as 
strong sensitivity to the mass of the heaviest element 
in the detector.
Several implications, including sizable recoil energy-dependent 
annual modulation, 
and improvements for future experiments are discussed.  
We show that the xenon experiments can improve on the PICO results, 
if they were to analyze their existing data over 
a larger range of recoil energies, i.e., $20$-$500$~keV\@. 
Intriguingly, CRESST has 
reported several events in the recoil energy range 
$45$-$100$~keV that, if interpreted as dark matter scattering, 
is compatible with $\delta \sim 200$~keV and an approximately 
weak scale cross section.
Future data from PICO and CRESST can test this speculation, while
xenon experiments could verify or refute this upon analyzing their
higher energy recoil data. 

\end{abstract}

\newpage

\section{Introduction}

At present, dark matter (DM) remains an uncharted component of our cosmos. Many experimental efforts are afoot to unmask non-gravitational interactions dark matter may have with known particles. Ongoing terrestrial direct detection searches for dark matter have excluded many well-motivated dark matter candidates. Nevertheless, dark matter may possess unusual characteristics that have allowed it to escape detection. This study focuses on the ``Inelastic Frontier'' -- dark matter scattering off nuclei into an excited state with a mass splitting broadly in the hundreds of keV range.  
While inelastic dark matter is by now well known \cite{Hall:1997ah,TuckerSmith:2001hy,TuckerSmith:2004jv,Finkbeiner:2007kk,Arina:2007tm,Cui:2009xq,Fox:2010bu,An:2011uq, Pospelov:2013nea,Dienes:2014via,Barello:2014uda}, much 
of the literature has focused on using the inelastic
transition to explain the DAMA/LIBRA annual modulation \cite{TuckerSmith:2001hy,TuckerSmith:2004jv,Chang:2008gd,Cheung:2009qd,Graham:2010ca,Barello:2014uda}.
Despite valiant model building attempts \cite{ArkaniHamed:2008qn,Alves:2009nf,Lisanti:2009am,Chang:2010en,Schwetz:2011xm,Weiner:2012gm,Barello:2014uda}, an inelastic
dark matter explanation is now extremely difficult to reconcile with
current data from several different experiments.  Hence, 
we do \emph{not} consider the DAMA/LIBRA annual modulation 
to be a signal of dark matter.  Instead, our focus is to 
consider the full range of inelastic splittings allowed by
kinematics, determining both the existing bounds, as well as 
improved bounds that could be obtained using already collected data,
reanalyzed at recoil energies well beyond the maximum recoil energy employed in some experiments' analyses.  

The predominant interaction of inelastic dark matter with Standard Model particles is mediated by an interaction $X_1 X_2 \mathcal{O}$, where $\mathcal{O}$ is an operator built from  standard model field(s).  The interaction could be dimension-4, e.g., interacting with the $Z$ or Higgs boson, or a higher dimensional interaction.
Assuming $X_1$ is the primary dark matter agent, the initial kinetic energy of the $X_1$-nuclear system must be greater than the mass difference ($\delta \equiv m_{\rm X_2} - m_{\rm X_1}$) between $X_1$ and $X_2$ in order for scattering to take place. The fact that \emph{only} DM with sufficiently large kinetic energy can scatter has two important consequences:
\begin{enumerate}
\item Because inelastic dark matter must impinge with sufficiently large kinetic energy to scatter with a direct detection target nucleus, the available kinematic phase space for DM-nuclear scattering is reduced, and the effective DM-nuclear scattering rate is suppressed. The amount of suppression will depend on what fraction of dark matter in the Galactic halo has enough kinetic energy to overcome the inelastic scattering energy threshold.
\item The minimum required energy for inelastic DM-nuclear collisions implies a minimum recoil energy in the detector, $E_{\rm R}^{\rm min}$. Traditional dark matter searches have optimized sensitivity to \emph{elastic} DM-nuclei collisions by focusing on the limit $E_{\rm R}^{\rm min} \rightarrow 0$, and pushing the observed recoil energy window as low as backgrounds and detector sensitivities allow, for example $1 \lesssim E_R~ \lesssim 30$~keV at LUX. For inelastic dark matter with a sizable mass splitting, a low maximum recoil energy reduces detector sensitivity. At best, a low maximum recoil energy will be sub-optimal for detecting inelastic dark matter. At worst, if the dark matter's \emph{minimum} inelastic recoil energy lies above the window of recoil energies, considered in a direct detection analysis (e.g. $E_{\rm R}^{\rm min} > 30$~keV at LUX), the experiment is {\em insensitive} to inelastic dark matter.  
\end{enumerate}
Both of these consequences are entirely a result of kinematics and hold regardless of the details of the inelastic dark matter model.  The bubble chamber experiment PICO, and earlier COUPP, is a notable exception in that it does not have an upper limit on the recoil energy to which it is sensitive.\footnote{In practice, all events with recoil energy $E_R\ltap 1$ MeV are expected to be accepted~\cite{privatecomm}.  Such a high cutoff gives PICO sensitivity to all of the parameter space we are interested in.}
As we will show, the overall implication is that current xenon-based dark matter experiments are insensitive to inelastic splittings $\delta \gtrsim 180$~keV, tungsten-based dark matter experiments are insensitive to $\delta \gtrsim 350$~keV\@, and bubble chamber experiments presently have the strongest constraints for $160 \ltap \delta\ltap 300$ keV\@.  By including larger nuclear recoil energies in future analyses, the combined reach of these experiments can be extended to $\delta \sim 550$~keV\@, and the iodine based-results can be improved upon by over an order of magnitude in much of the parameter space, just using present xenon exposures. 
Moreover, our study reveals that because tungsten is the heaviest element currently employed in dark matter experiments, experiments like CRESST can probe the largest inelastic dark matter mass splitting.\footnote{The heavier element thallium ($A=205$) is used as a doping agent in the DAMA~\cite{Bernabei:2010mq} and KIMS~\cite{Kim:2012rza} experiments, with $O(10^{-3})$ concentrations.  This impurity is too small to provide a meaningful bound but has been investigated in the past as a possible inelastic explanation of the DAMA excess~\cite{Chang:2010pr}.}

There are now many examples of well-motivated dark matter models that predict sizable inelastic mass splittings, in this work we will consider three examples. A supersymmetric example is Higgsino dark matter with $\mathcal{O}(100)$~keV mass splittings between the Higgsino states induced by mixing with very heavy, or carefully tuned, neutralinos.  In this case the coupling to the nucleus is through $Z$ exchange.  For our second example, we consider DM-nucleus interactions that take place through photon exchange. Specifically, we will focus on magnetic inelastic dark matter (MIDM) \cite{Chang:2010en,Weiner:2012gm}, where the DM-photon coupling is a transition dipole operator, between states split by $\mathcal{O}(100)$~keV\@.  In our third example, the state exchanged during direct detection is a dark photon of mass $\sim 0.1-10$ GeV\@; here the dark sector inelastic splitting arises from coupling to the scalar that makes the dark photon massive.  Further details on these models will be given in Sec.~\ref{sec:models}.

The setup of the rest of this paper is as follows.  In Sec.~\ref{sec:sbasics}, we explore the kinematics of inelastic scattering, particularly studying how detector nuclear recoil energies and velocity phase space affect searches for dark matter with a large inelastic mass splitting.  In Sec.~\ref{sec:rate}, we examine the dark matter-nuclear scattering rate at large inelastic mass splittings and recoil energies for xenon and tungsten targets.  Sec.~\ref{sec:prosp} finds how existing and prospective dark matter direct detection analyses can bound inelastic dark matter.  In Sec.~\ref{sec:models}, we explore a number of dark matter models that give rise to a large inelastic mass splitting.  In Sec.~\ref{sec:cresstwimp}, we note that four events recently observed in the $E_{R}=30$-$120$\,keV recoil energy band at CRESST would be consistent with 1 TeV WIMP-like dark matter with a $\sim 200$\,keV mass splitting.  In Sec.~\ref{sec:conclusions} we conclude.

\section{Inelastic Kinematics}
\label{sec:sbasics}

\begin{figure}
\centering
\includegraphics[width=0.49\textwidth]{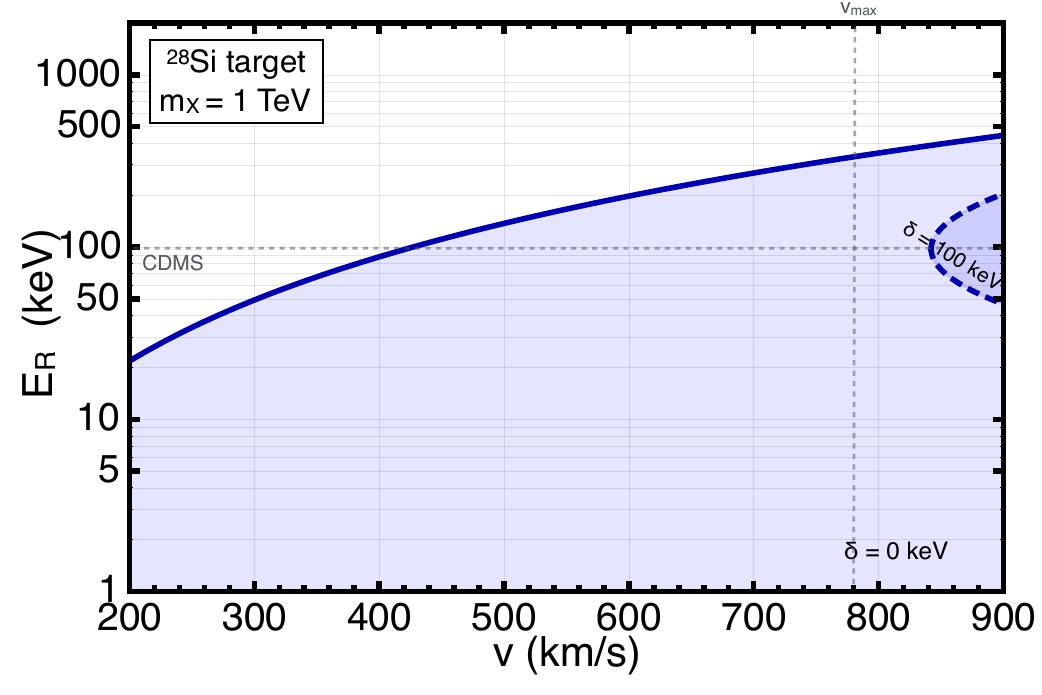}
\includegraphics[width=0.49\textwidth]{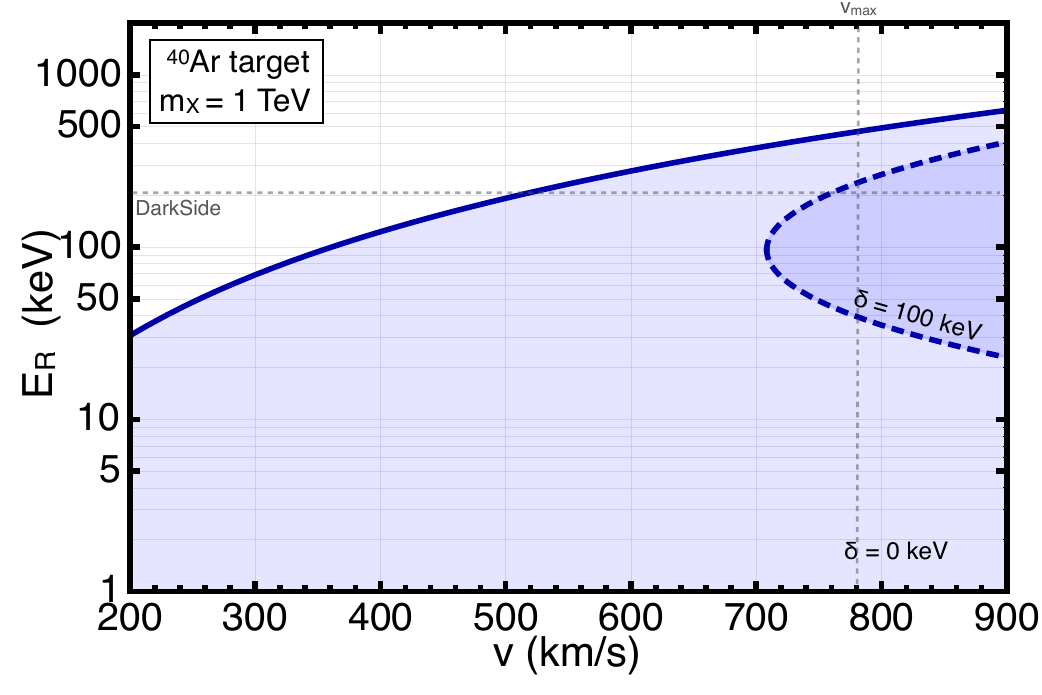}
\includegraphics[width=0.49\textwidth]{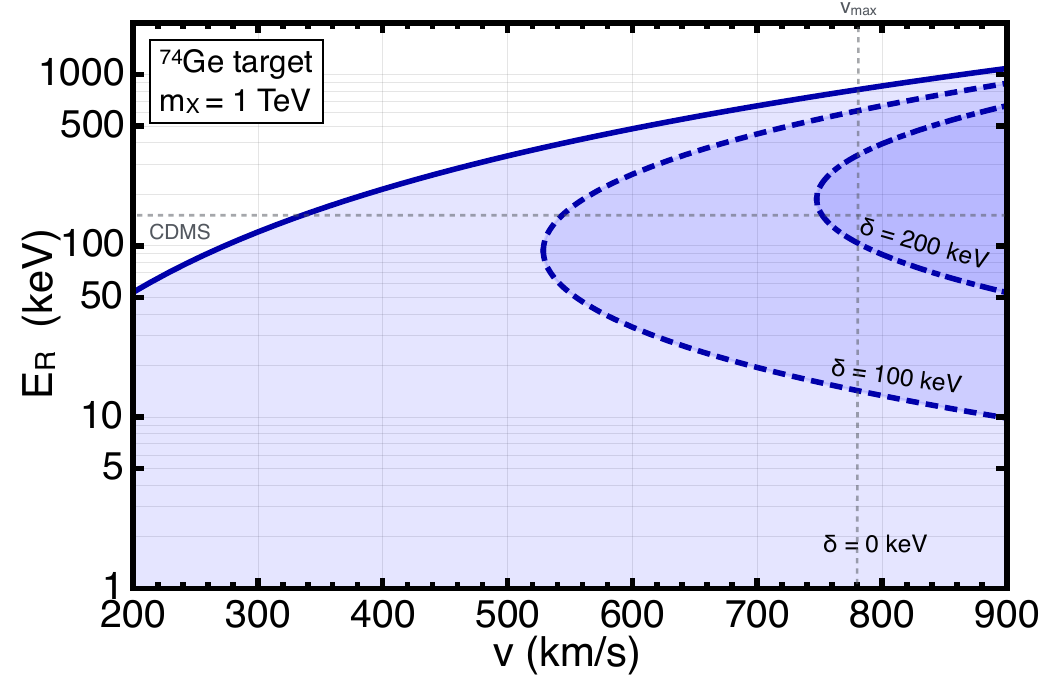}
\includegraphics[width=0.49\textwidth]{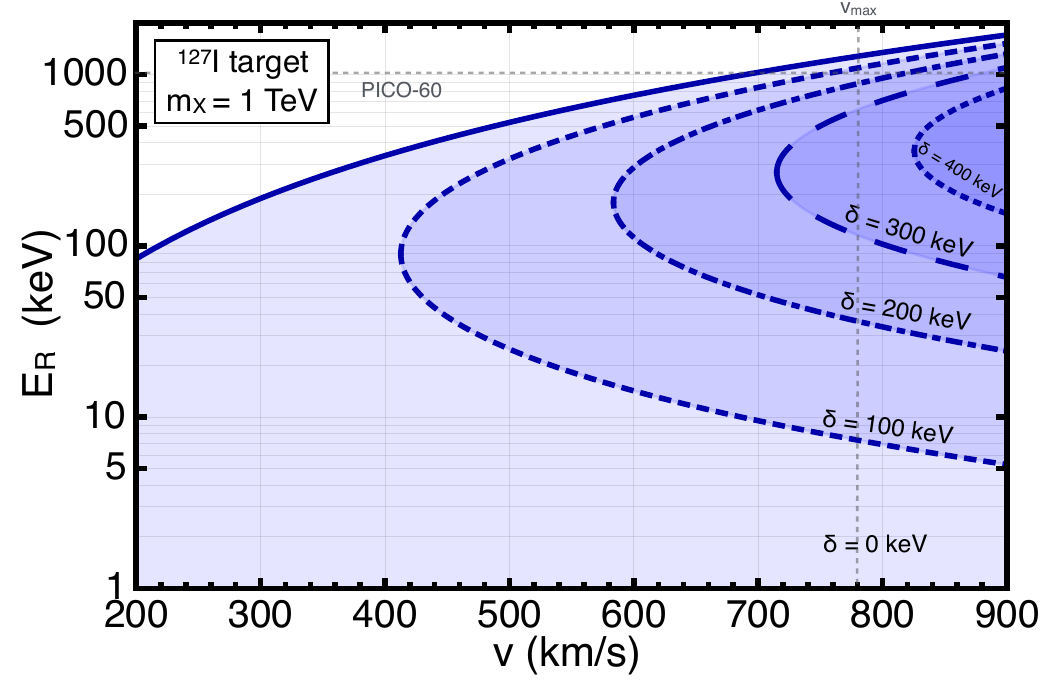}
\includegraphics[width=0.49\textwidth]{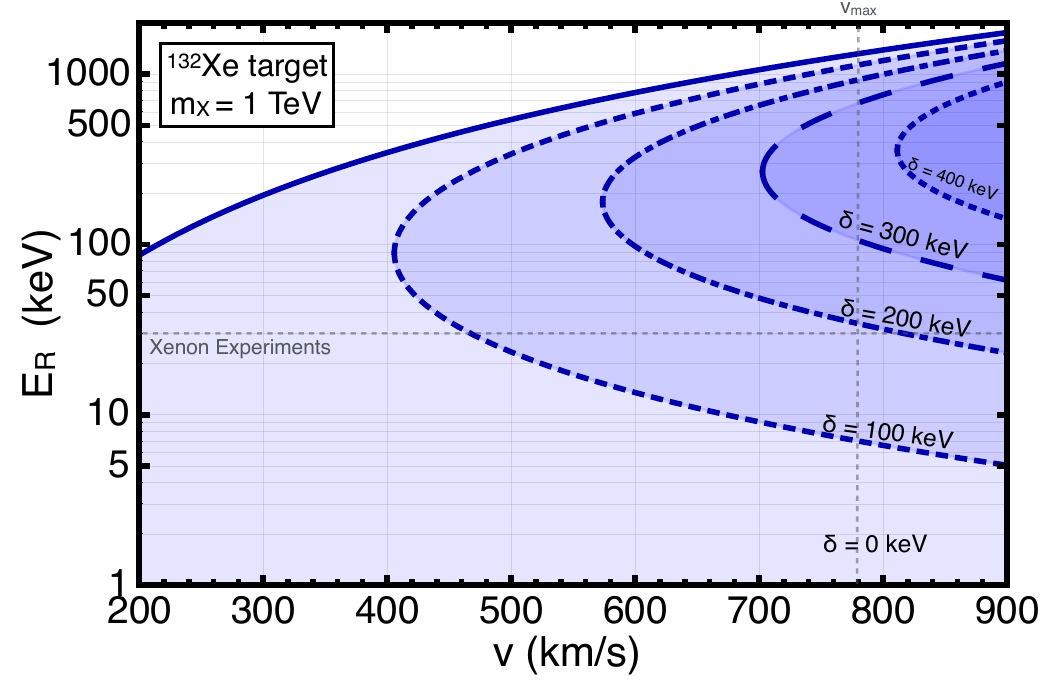}
\includegraphics[width=0.49\textwidth]{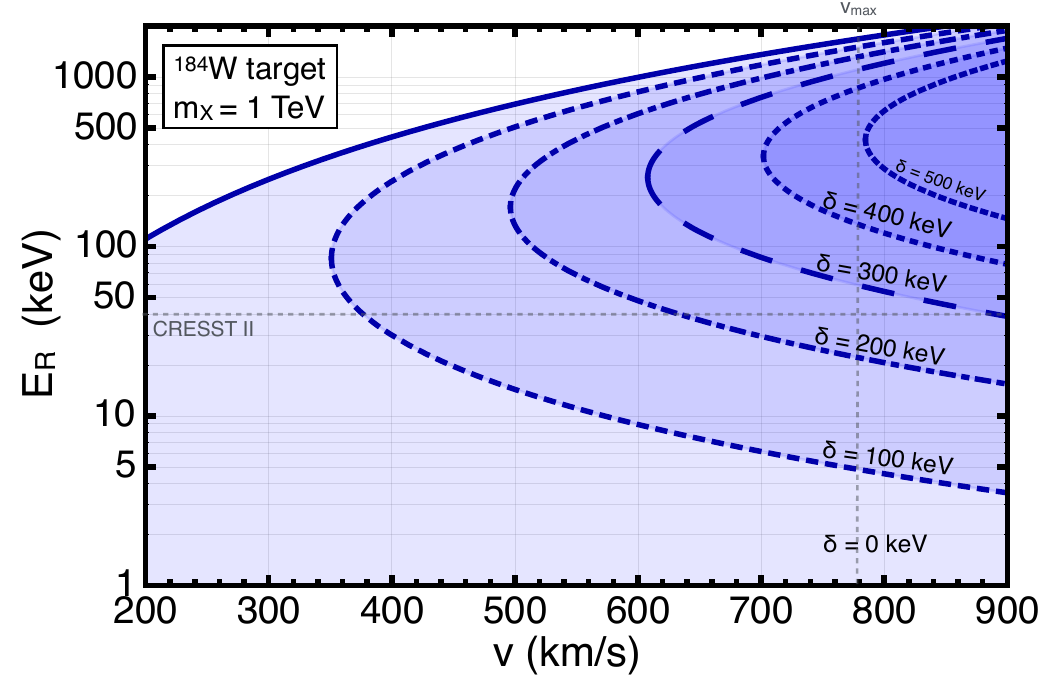}
\caption{The shaded region is the available range of recoil energies on a nuclear target, for a given DM mass splitting and incoming DM speed in the laboratory frame. The contours indicate mass splitting $\delta = 0$ in solid, $\delta = 100, 200, 300, 400, 500$~keV in dashed, dotted-dashed, dotted, long-dashed, and fine dotted, respectively. The dashed grey horizontal lines indicate the maximum recoil energy windows used by collaborations including CDMS \cite{Ahmed:2010hw,Agnese:2013rvf}, DarkSide \cite{Agnes:2015lwe}, PICO-60 \cite{Amole:2015pla}, Xenon Experiments (LUX \cite{Akerib:2015rjg}, PandaX II \cite{Tan:2016zwf}, XENON100 \cite{Aprile:2012nq}), and CRESST II \cite{Angloher:2015ewa}. Note that the \emph{maximum} incoming terrestrial dark matter speed is expected to be $780^{+54}_{-41} ~{\rm km/s}$, Ref.~\cite{Piffl:2013mla}.}
\label{fig:basics}
\end{figure}

To begin, we examine kinematic properties of inelastic dark matter (see $e.g.$ \cite{TuckerSmith:2001hy,Petriello:2008jj,Chang:2008gd,Cui:2009xq,Chang:2010pr} for prior discussion). Here we will scrutinize the impact of dark matter inelastic mass splitting $\delta$ and dark matter velocity $v$ on the spectrum of expected nuclear recoil event energies at fixed target dark matter searches.
The kinetic energy of a dark matter particle is $E_0 = \frac{1}{2} m_X v^2$, where the velocity $v$ is in the laboratory frame. In the non-relativistic limit appropriate for dark matter scattering, the recoil energy is 
\begin{eqnarray}
E_R &=& \frac{\mu}{m_N} \left[
        \left( \mu v^2 \cos^2\tlab - \delta \right) 
\pm (\mu v^2 \cos^2\tlab)^{1/2} 
        \left( \mu v^2 \cos^2\tlab - 2 \delta \right)^{1/2} \right] \nonumber
        \;  , 
\end{eqnarray}
where $\tlab$ is the scattering angle in the laboratory frame, and $\mu$ is the reduced mass of dark matter and the nucleus.  
We illustrate the kinematically allowed space of recoil energies 
as a function of velocity in Fig.~\ref{fig:basics}. 
The contours correspond to maximal scattering angle, 
i.e., $\cos^2\tlab = 1$.  In this figure, we show several 
target elements for a fixed dark matter mass $m_X = 1$\,TeV and
several different inelastic splittings.

Notice that there is both a lower bound \emph{and} an upper
bound on the recoil energy, assuming a maximum incoming
terrestrial dark matter speed.  Unsurprisingly, the maximum
recoil energy asymptotes to the maximum recoil energy for 
elastic scattering once $E_R \gg \delta$.  The minimum
velocity to scatter at \emph{any} recoil energy
is determined by the apex of the parabola 
\begin{eqnarray}
v^{\rm apex}_{\rm min} &=& \sqrt{\frac{2 \delta}{\mu}} \\
E_R(v^{\rm apex}_{\rm min}) &=& \frac{\mu}{m_N} \delta \, .
\end{eqnarray}

\begin{figure}[t!]
\centering
\includegraphics[width=0.48\textwidth]{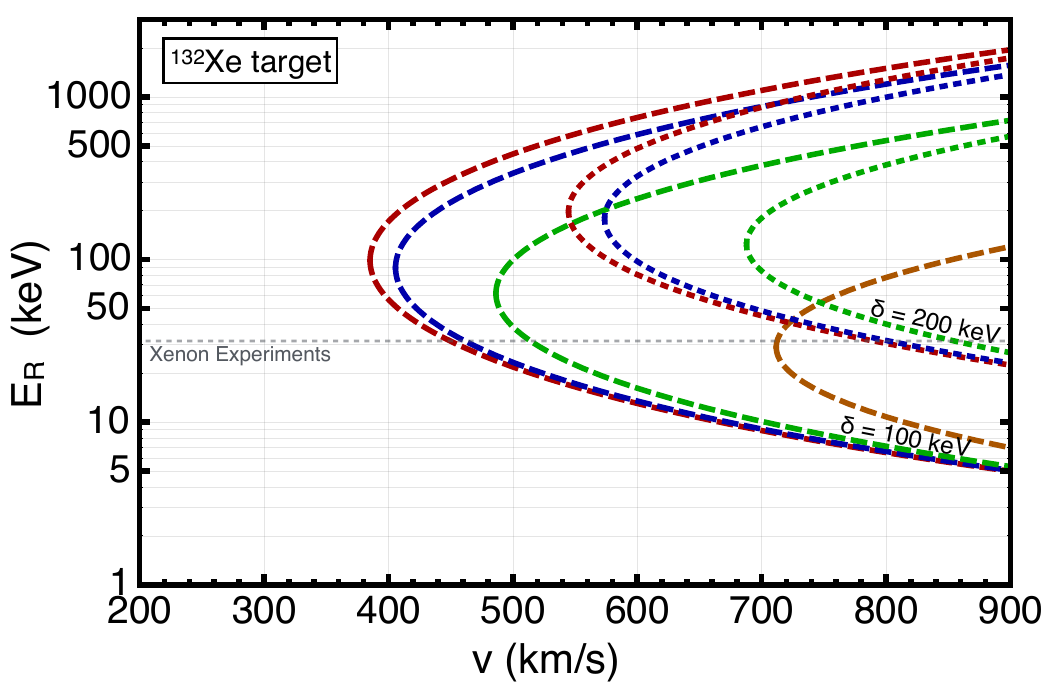}
\includegraphics[width=0.48\textwidth]{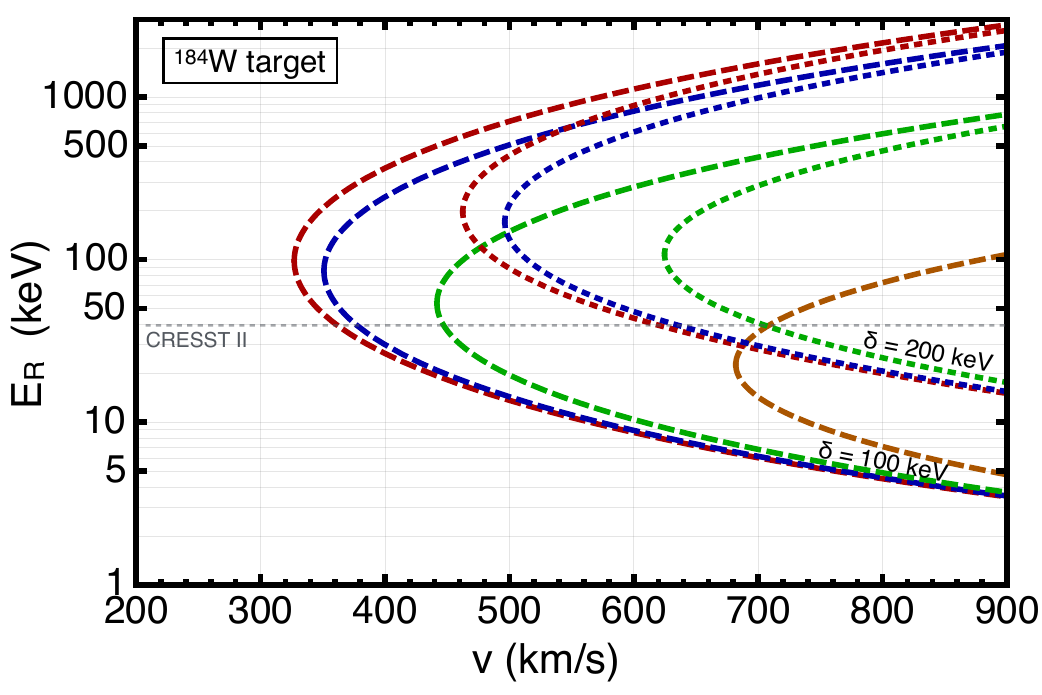}
\caption{The energy recoil boundaries for inelastic DM with splitting $\delta = 100$~keV (dashed) and $200$~keV (dotted) scattering off xenon and tungsten. From right to left, the orange, green, blue, and red curves denote available scattering phase space for $m_X = 0.05,0.2,1,10 ~{\rm TeV}$ dark matter, respectively. Dark matter masses $m_X > 10 ~{\rm TeV}$ are nearly indistinguishable from the $m_X=10$~TeV curve due to the reduced mass $\mu \simeq m_N$. As in Fig.~\ref{fig:basics}, horizontal lines indicate the maximum recoil energies of CRESST II, LUX, PandaX II, and XENON100.}
\label{fig:basics_2}
 \end{figure}

The general result as a function of $E_R$ is
\begin{equation}
v_{\rm min} = \frac{1}{\sqrt{2 E_R m_N}}
              \left( \frac{E_R\, m_N}{\mu} + \delta \right) \, ,
\label{eq:vmin}
\end{equation}
where this expression is valid up to corrections of 
$\mathcal{O}(E_0/m_X, \delta/m_X)$, which are negligible for 
fixed-target terrestrial experiments and $m_X \gg {\rm GeV}$. 
The mass dependence of the kinematics is illustrated in 
Fig.~\ref{fig:basics_2}.

The simple expressions for the apex make it easy 
to qualitatively understand how the kinematical range
shifts with respect to the dark matter mass, 
the target mass, and the inelastic splitting. For example, in the case of heavy dark matter $m_X \gg m_N$, 
\begin{equation}
v^{\rm apex}_{\rm min} \; \simeq \; \sqrt{\frac{2 \delta}{m_N}} \qquad 
E_R(v^{\rm apex}_{\rm min}) \; \simeq \; \delta \qquad (m_X \gg m_N) \, , 
\end{equation}
the minimum velocity scales as $1/\sqrt{m_N}$, which makes
it clear why argon experiments have essentially no sensitivity to 
the inelastic frontier.  Also, the typical recoil energy
is determined just by $\delta$, independently of the
dark matter mass and the target nucleus.  Therefore,
experiments that employ a maximum $E_R$ that is
less than $\delta$ are severely limiting 
their sensitivity.  This is illustrated in the figure
by the maximum $E_R$ that existing analyses from 
LUX, PandaX, and XENON100 use to set their bounds.
By only accepting events with recoil energies smaller 
than $\sim 50$~keV, these analyses are necessarily restricted to 
inelastic mass splittings less than $200$~keV.

In the case of light dark matter $m_X \ll m_N$,
\begin{equation}
v^{\rm apex}_{\rm min} \; \simeq \; \sqrt{\frac{2 \delta}{m_X}} \qquad 
E_R(v^{\rm apex}_{\rm min}) \; \simeq \; \frac{m_X}{m_N} \delta \qquad (m_X \ll m_N) \, , 
\end{equation}
the minimum velocity to scatter is larger than in the
case of heavy dark matter by a factor of $\sqrt{m_N/m_X}$.
Holding $\delta$ fixed, for even a modest reduction of $m_X$,
the minimum velocity to scatter will exceed the maximum
incoming terrestrial dark matter speed, and thus there is 
no sensitivity.  If $\delta$ and $m_X$ are reduced simultaneously, the typical recoil energy
also is reduced, and this can place it within 
the window of recoil energies considered by existing experimental analyses.
This is not surprising; by reducing $\delta,\,m_X,$ and, by extension, $E_R$, we approach 
the limit of elastic scattering and experimental analyses optimized 
for elastic scattering will be able to set bounds
once $\delta$ is small compared with the kinetic
energy of the dark matter.  This is illustrated by
the \emph{absence} of orange $m_X = 50$ GeV, $\delta \sim 200$ keV contours in Fig.~\ref{fig:basics_2}: $i.e.$ inelastic sensitivity to 50 GeV mass dark matter would not be greatly improved by searching at higher recoil energies, since even at maximum incoming velocity, such light dark matter will not surmount the $\delta \sim 200 ~{\rm keV}$ threshold. 

It is already clear from Figs.~\ref{fig:basics} and \ref{fig:basics_2} 
that the ``frontier'' of inelastic dark matter is when the dark matter
is heavy, $\mu \simeq m_N$, and can be tested by experiments that have 
heavy mass elements: iodine (PICO), xenon (LUX, PandaX, XENON100), 
and tungsten (CRESST).  For the purposes of illustration,
we use $m_X = 1\, \tev, 10\, \tev$ as benchmarks for much of
the remainder of the paper.    
We also emphasize that since the inelastic frontier
occurs at relatively large velocities and recoil energies,
we anticipate substantial astrophysical and experimental uncertainties 
in the scattering rates and comparisons between experiments,
that we will elaborate on more below. 

\section{Inelastic direct detection at high recoil energies}
\label{sec:rate}

For a given recoil energy $E_R$, the rate at which DM scatters in a detector with $N_{\rm T}$ scattering targets per unit mass is given by 
\begin{equation}
\frac{d R}{d E_{\rm R}} \; = \; n_{X}  N_{\rm T} \int_{v_{\rm min} }^{v_{\rm max}} v~f(\vec{v},\vec{v}_{\rm e})~\frac{d \sigma}{d E_{\rm R}}~ d^3v,
\label{eq:rateER}
\end{equation}
where $n_{X} = \rho_{X}/m_{X} $ is the local dark matter number density, $\vec{v}$ is dark matter's and $\vec{v}_{\rm e}$ is Earth's velocity in the Galactic rest frame, $v_{\rm min}$ is the minimum dark matter speed in the detector's rest frame (Eq.~(\ref{eq:vmin})), $v_{\rm max}$ is the maximum incoming DM speed in the earth's frame of reference, and $\dd \sigma / \dd E_{\rm R}$ is the differential cross-section for dark matter scattering off a nucleus. In terms of the parameter space illustrated in Figs.~\ref{fig:basics},~\ref{fig:basics_2}, Eq.~\eqref{eq:rateER} corresponds to picking a value on the vertical axis and integrating along the horizontal axis starting at the apex of a contour and ending at $v_{esc}$. The total scattering rate over all recoil energies is the integral of Eq.~\eqref{eq:rateER} over the relevant experiment's energy window. To understand all of the physics in Eq.~\eqref{eq:rateER} and how it is affected by inelasticity, we discuss the two terms in the integrand separately.

\subsubsection*{Velocity distribution}

Figures~\ref{fig:basics},~\ref{fig:basics_2} have demonstrated that the Milky Way DM velocity distribution play a major role in determining what DM phase space is available. For this study, we utilize a Maxwellian distribution with a sharp cutoff at $v_{esc}$ and incorporating the relative velocity between the earth and the sun\footnote{Using a velocity distribution where $f(v)\rightarrow 0$ smoothly at $v_{esc}$, as in Ref.~\cite{Lisanti:2010qx}, does not qualitatively change our results.}\,
\begin{equation}
f(\vec{v},\vec{v}_{\rm e}) \; = \; \frac{e^{-(v^2+v_{\rm e}^2+2v v_{\rm e}{\rm cos}~\theta)/v_0^2}}{N(v_0,v_{\rm esc})},
\end{equation}
where $\theta$ is the angle between Earth's ($\vec{v}_{\rm e}$) and dark matter's ($\vec{v}$) velocity in the Galactic rest frame. The 
factor $N(v_0,v_{\rm esc})$ normalizes the velocity distribution so that $\int_0^{v_{\rm max}} f( \vec{v} , \vec{v}_{\rm e} ) ~\dd^3 v=1$. For a galaxy with escape speed $v_{\rm esc}$ and characteristic dark matter speed $v_0$, the escape speed and maximum velocity in the earth's frame  ($v_{max}$) are related by:
\begin{equation}
v^2_{esc} \; = \; v^2_{max} + v^2_e + 2\, v_{max}\, v_e\, \cos{\theta}~.
\label{eq:angrest}
\end{equation}
The normalization factor is
\begin{equation}
N(v_0,v_{\rm esc}) \; = \; \pi^{3/2} v_0^3 \left( {\rm erf} \left(\frac{v_{\rm esc}}{v_0} \right)-\frac{2 v_{\rm esc}}{\pi^{1/2}v_0} e^{-v_{\rm esc}^2/v_0^2} \right).
\end{equation}
For the earth's velocity around the sun, we take the expression ($e.g.$ \cite{Lewin:1995rx})
\begin{equation}
v_{\rm e} \; = \; \left[ 232 + 15\, \cos \left( \frac{2 \pi(t-152~{\rm days})}{365~{\rm days}} \right) \right] ~{\rm km/s},
\label{eq:ve}
\end{equation}
where $0<t<365$ denotes days of the year, beginning with $t=0$ on January 1st.  This expression gives a reasonable approximation to more accurate determinations \cite{Lee:2013xxa,McCabe:2013kea}.  Within this approximation the maximum lab frame speed ($v= 247~ {\rm km/s}$) is attained on June 2nd, $t=152$. The relative Earth-Sun velocity results in an annual modulation of the scattering rate.

For finite $v_{esc}$, the velocity integral depends on the relative order of $v_{\rm min}$ and $v_{\rm esc} -v_{\rm e}$~\cite{Lewin:1995rx}:
\begin{equation}
\int_{v_{\rm min}}^{v_{\rm max}} \, d^3 v \; = \; 2 \pi \left\{ \begin{array}{cc} \int_{v_{\rm min}}^{v_{\rm esc} -v_{\rm e}}~v^2 dv  \int_{-1}^{1}~d\!\cos{\theta} +\int_{v_{\rm esc}-v_{\rm e}}^{v_{\rm esc}+ v_{\rm e}}~v^2 dv \int_{-1}^{c_*} d\!\cos{\theta}  & v_{\rm min} < v_{\rm esc}-v_{\rm e} \\
& \\
\int_{v_{\rm min}}^{v_{\rm esc}+ v_{\rm e}}~v^2 dv \int_{-1}^{c_*} d\!\cos{\theta}  & v_{\rm esc}-v_{\rm e} < v_{\rm min} < v_{\rm esc} + v_{\rm e}  \\
\end{array}\right.
\label{eq:fullintegral}
\end{equation}
where $c_* = {\rm cos}~\theta_* = (v_{\rm esc}^2 -v^2-v_{\rm e}^2)/(2\, v\, v_{\rm e})$ is the minimum angle consistent for scattering when $v_{\rm esc}-v_{\rm e}\leq v \leq v_{\rm esc}+v_{\rm e}$. Note that the limits of the velocity integrals will change depending on the time of year due to the time dependence in $v_e$. For inelastic scenarios with high $v_{min}$, this can potentially lead to dramatic features, such as scattering that only occurs during some short interval around June 2nd. We will say more about the modulation of highly inelastic scenarios in Sec. \ref{sec:prosp}.\\

\subsubsection*{Differential cross section}
The differential cross section $d\sigma/dE_R$ contains the details of how DM interacts with SM fields. To begin our study of highly inelastic DM, we take DM-nuclei scattering to be spin- and energy-independent; for this scenario, $d\sigma/dE_R$ is customarily given as
\begin{equation}
\frac{d\sigma}{d E_{\rm R}} \; = \; \frac{\sigma_{\rm n}}{v^2}\frac{m_{\rm N}}{2 \mu_{\rm n}^2} \frac{\left[Z f_p +(A-Z)f_n \right]^2}{f_n^2} F^2(E_{\rm R}),
\label{eq:dsnde}
\end{equation}
where $\sigma_{\rm n}$ and $\mu_{\rm n}$ are the dark matter-nucleon scattering cross-section and reduced mass, $A$ and $Z$ are the nuclear atomic mass and number, and $f_p(f_n)$ encapsulate the DM-proton (DM-neutron) effective couplings. All of the energy dependence lies in $F^2(E_{\rm R})$, the nuclear form factor that characterizes how coherently dark matter scatters off the nucleus.

Detailed recoil energy dependent form factors have been calculated using nuclear physics models for several relevant DM scattering elements and isotopes. Whenever possible, we use the results of the most recent calculations, notably Ref.~\cite{Vietze:2014vsa} calculate form factors for xenon. For elements/isotopes where calculations are not publicly available, such as tungsten and iodine, we will use the Helm form factor. The form factors (either Helm or from dedicated nuclear calculations) suppress higher-energy scattering events which probe the sub-structure of the nuclei. As inelastic scattering involves large recoil energies, form factor suppression will play a much larger role than in elastic scattering. Additionally, the (spin-independent) form factors have several `zeros', recoil energies corresponding to momentum exchanges where the scattering contributions from different nucleons destructively interfere. The specifics of the form factors we use can be found in appendix~\ref{app:formfactor}. \\

\begin{figure}[t!]
\center
\includegraphics[scale=1.1]{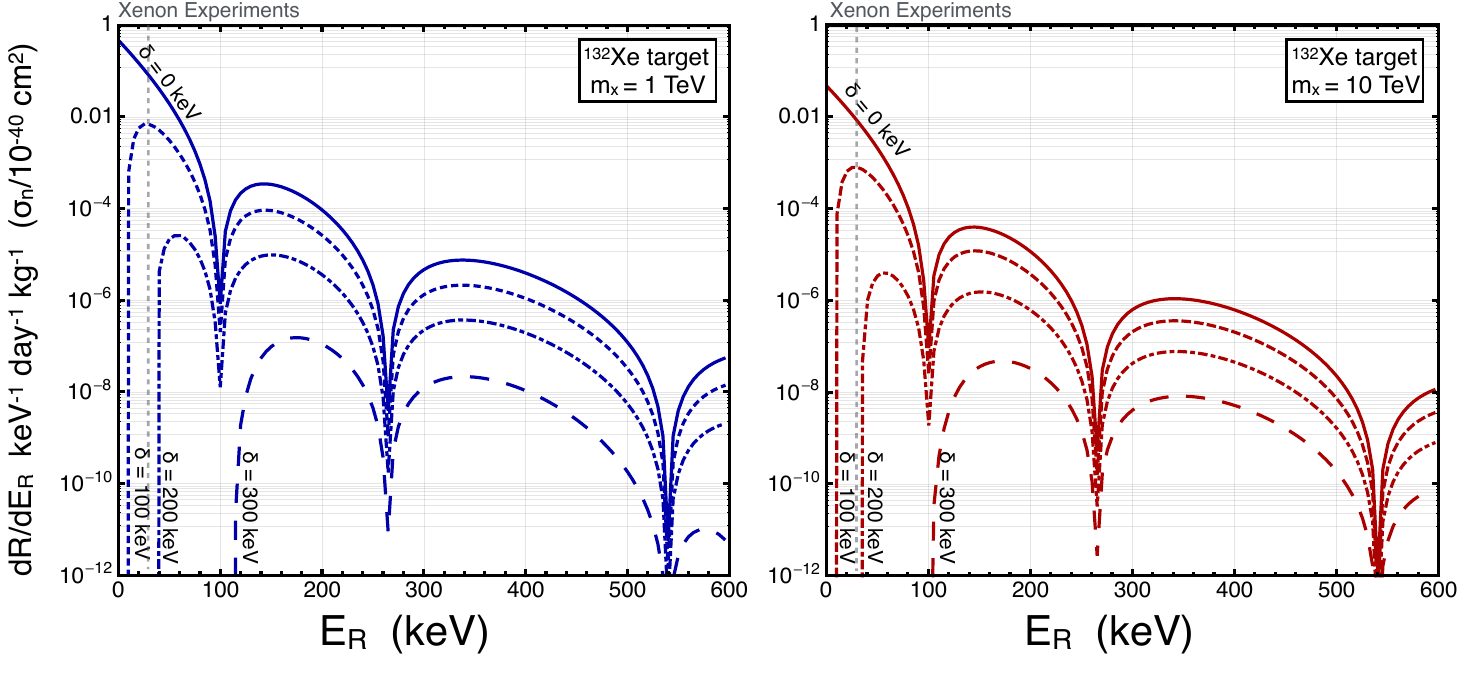}
\caption{Rate for dark matter nucleon scattering assuming a DM-nucleon cross-section $\sigma_{\rm n} = 10^{-40} \; {\rm cm}^2$ and a target made purely of $^{132}$Xe. Blue (red) lines indicate $d\sigma/dE_R$ for $m_X = 1 {\rm ~ TeV}$ ($=10$~TeV) inelastic dark matter, with $\delta = 0, 100, 200, 300, 400$~keV mass splittings between dark matter states, as indicated. The vertical line marks the maximum recoil energy considered by LUX in \cite{Akerib:2015rjg}.}
\label{fig:Ereach}
\end{figure}

\begin{figure}[t!]
\center
\includegraphics[scale=1.1]{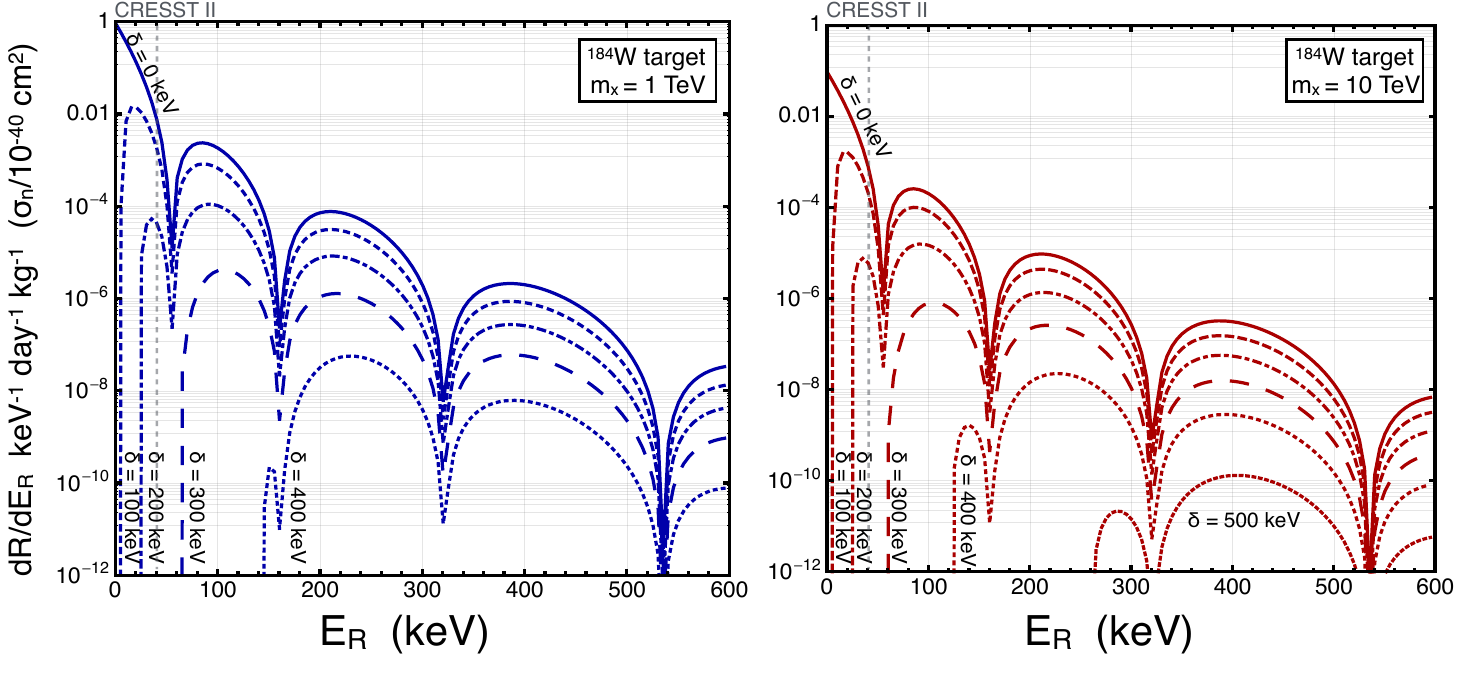}
\caption{Rate for dark matter nucleon scattering per kg per day and per keV of nuclear recoil energy, assuming a DM-nucleon cross-section $\sigma_{\rm n} = 10^{-40} \; {\rm cm}^2$ and a target made purely of $^{184}$W. Blue (red) lines indicate $d\sigma/dE_R$ for $m_X = 1 {\rm ~ TeV}$ ($10~{\rm TeV}$) inelastic dark matter, with $\delta = 0, 100, 200, 300, 400, 500$~keV mass splittings between dark matter states, as indicated. The vertical line marks the maximum recoil energy considered by CRESST II in Ref.~\cite{Angloher:2015ewa}.}
\label{fig:Ereach2}
\end{figure}

We are now ready to examine the recoil rate as a function of inelastic mass splitting. We will look at two different DM masses, $1\, \tev$ and $10\, \tev$, colliding with two different nuclear targets, xenon ($A = 132$), and tungsten ($A = 184$). The DM mass and nuclear parameters completely specify the spin-independent scattering $d\sigma/dE_R$ in the limit of equal DM-proton and DM-neutron couplings ($f_n = f_p$), up to an overall scaling by $\sigma_{\rm n}$. For both masses and targets, we take $v_{esc} = 533~ \kms$, average velocity $v_0 = 220~ \kms$  and pick spring/autumn so that $v_e = 232~{\rm km/s}$, and assume a DM density of $0.3\, \gev/{\rm cm}^3$. The only remaining input is the DM mass splitting $\delta$, which sets the minimum scattering velocity. 

As a first step, we fix $\sigma_{\rm n} \sim 10^{-40} \; {\rm cm}^2$, then plot the event rate (in events per keV-day-kg) of DM-nuclear scattering as a function of recoil energy; the results for a xenon target are shown in Fig.~\ref{fig:Ereach} and a tungsten target in Fig.~\ref{fig:Ereach2}.

These plots give insight into how $E_R$ and $\delta$ drive searches for highly inelastic dark matter. The trends shown in Fig.~\ref{fig:basics} become apparent in two ways: i.) as $\delta$ is increased, the phase space available for scattering shrinks, and the event rates decline, and ii.) the optimum recoil energy to look for a particular DM mass and splitting also increases with $\delta$. The sharp dips in the recoil rate shown in Figs. \ref{fig:Ereach} and \ref{fig:Ereach2} are a result of the nuclear form factor zeros.

Comparing the left and right panels in Fig.~\ref{fig:Ereach} or either panel in Fig.~\ref{fig:Ereach}  with the corresponding panel in Fig.~\ref{fig:Ereach2}, we can also see how the kinematic trends illustrated in Fig.~\ref{fig:basics_2} percolate through the full rate calculation.  For fixed DM mass, the rate of DM scattering off tungsten at large $\delta$ exceeds that of DM scattering off xenon.  As one specific example, in Fig.~\ref{fig:Ereach2} we see that the larger reduced mass in collisions of 10 TeV DM with tungsten means splittings $\delta = 500$~keV are accessible. This shows that, when setting constraints on highly inelastic DM, heavy nuclei are more effective.  Another way to see this is to note that for $m_X \gg m_N$, $v_{min}$ scales as $\frac{1} {\sqrt{m_N}},$ and thus the minimum velocity to scatter in tungsten is $85\%$ of that in xenon.

\begin{table}[t]
   \centering
   \begin{tabular}{@{} lccccc @{}} 
      \toprule 
        Experiment   & Exposure [tonne-days] & Energy range $[\mathrm{keV}_{nr}]$
& Candidate DM events & Refs.\\
      \midrule
    PICO            & 1.3       & 7-20--$\mathcal{O}(1)\,\mathrm{MeV}$ &0           & \cite{Amole:2015pla}\\
    LUX              & 14        & 1--30                            & 0           & \cite{Akerib:2015rjg} \\
    PandaX        & 33         & 1--30		             & 1           &\cite{Tan:2016zwf}\\
    CRESST      & 0.052   & $30$--$120$                   &          4  & \cite{Angloher:2015ewa} \\
      \bottomrule
   \end{tabular}
   \caption{Summary of experimental results.  For CRESST, the
upper bound on their signal region was $40$~keV, but all events up to 
$120$~keV were shown in Ref.~\cite{Angloher:2015ewa}, see
Sec.~\ref{sec:cresstwimp}. For PICO the lower bound was not
constant over the course of the data taking, but varied from 
$\sim 7$ to $\sim 20$~keV.}
   \label{tab:experimentalnumbers}
\end{table}

\section{Inelastic dark matter in existing data: rates and modulation}
\label{sec:prosp}

In prior sections we have explored the kinematics of DM inelastic scattering. We now explore what sensitivity could be attained with searches including higher nuclear recoil energy events, specifically for DM with a mass $m_X \gtrsim m_N$.  Our starting point is Eq.~(\ref{eq:rateER}); for a given DM mass/splitting, target element, and working in the limit of $f_p = f_n$, the only undetermined quantity is the overall scaling $\sigma_n$, which we constrain using results from several direct detection experiments. 
The experiments we consider are listed in Table~\ref{tab:experimentalnumbers}. The LUX collaboration's most recent published result found 0 nuclear recoil events consistent with a DM signal in a $1-30\, \kev$ recoil energy window after $1.4 \times 10^4$ kg days of running \cite{Akerib:2015rjg}. The PandaX II experiment has observed only 1 nuclear recoil event inside a $1-30\, \kev$ recoil energy window for $3.3 \times 10^4$ kg days of exposure \cite{Tan:2016zwf}. This PandaX result matches a recently reported bound from a similar LUX exposure \cite{LUXtalk}. For tungsten, we use some of the latest results from CRESST II: 4 signal-like events in a $30$-$120$~keV window after $52.2$~kg~days.\footnote{The CRESST II analysis strictly considers a recoil energy band $E_R = 0.3-40$~keV that had a large contamination from backgrounds at the low end of this range.  Since CRESST has shown their recoil energy data out to $120$~keV \cite{Angloher:2015ewa}, we employ the energy window $30$-$120$~keV\@.}  In the case of PICO \cite{Amole:2015pla} (iodine), events with recoil energies between $\sim 7-10^3$~keV were accepted over $\sim 1300$ kg days. For convenience, these integrated exposures and recoil energy bands are presented in Table \ref{tab:experimentalnumbers}. Assuming only statistical uncertainties and keeping the volume and energy intervals of the experiments fixed, the $90\%$ confidence level bound is $2.3$ events if no events are observed, $3.9$ events if one events is observed, or $6.7$ events if four events are observed.\footnote{Technically, LUX and PandaX employ profile likelihood methods to address their backgrounds, rather than simple cut-and-count.}

\begin{figure}[t!]
\center
\includegraphics[width=0.48\textwidth]{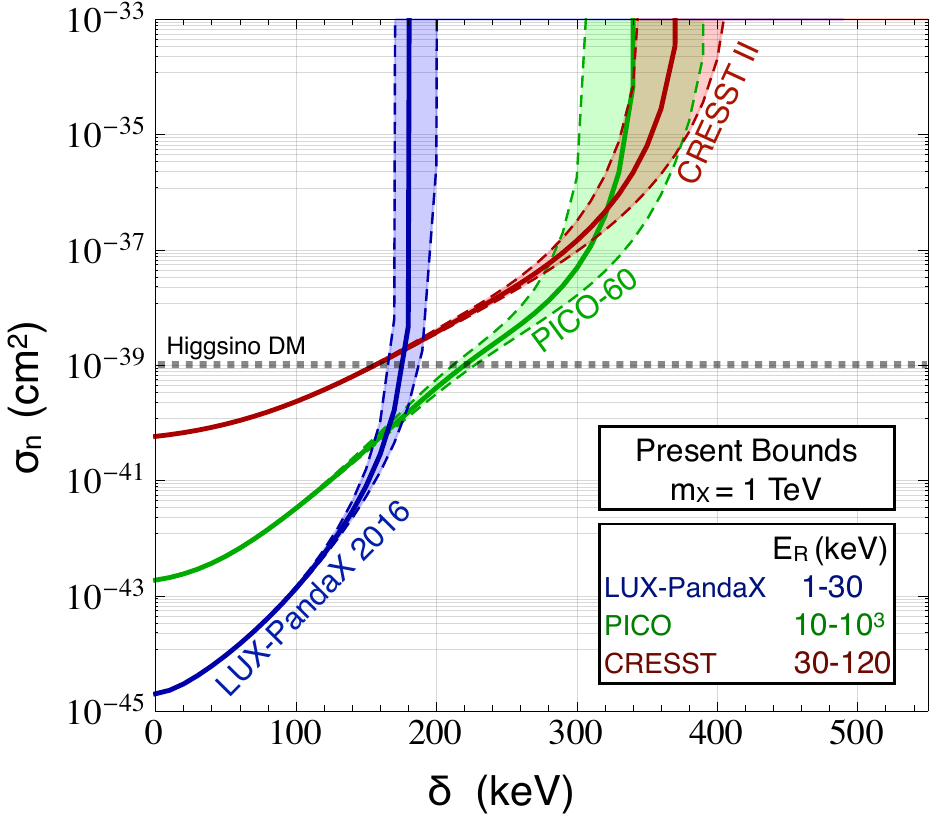}
\includegraphics[width=0.48\textwidth]{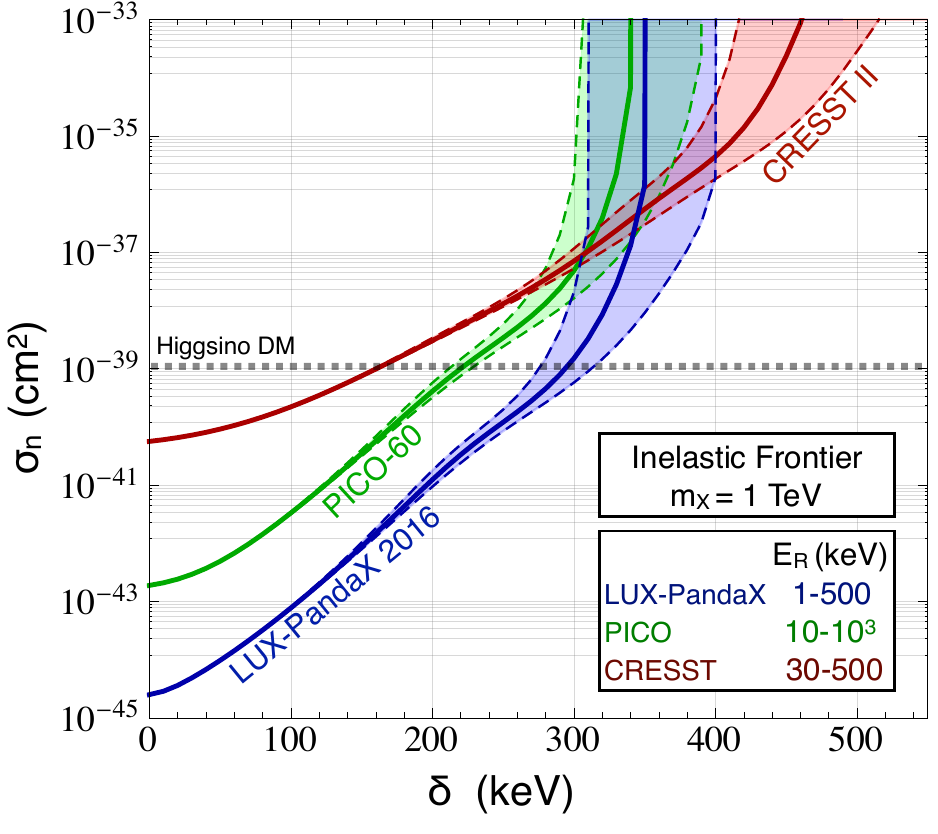}
\includegraphics[width=0.49\textwidth]{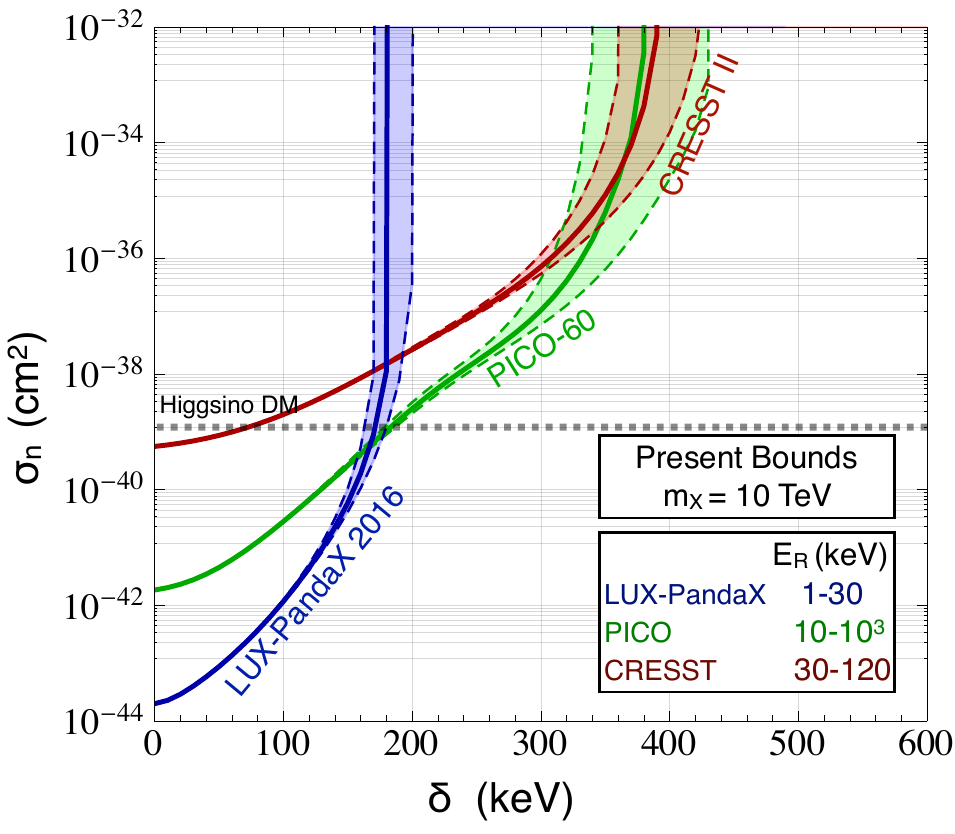}
\includegraphics[width=0.49\textwidth]{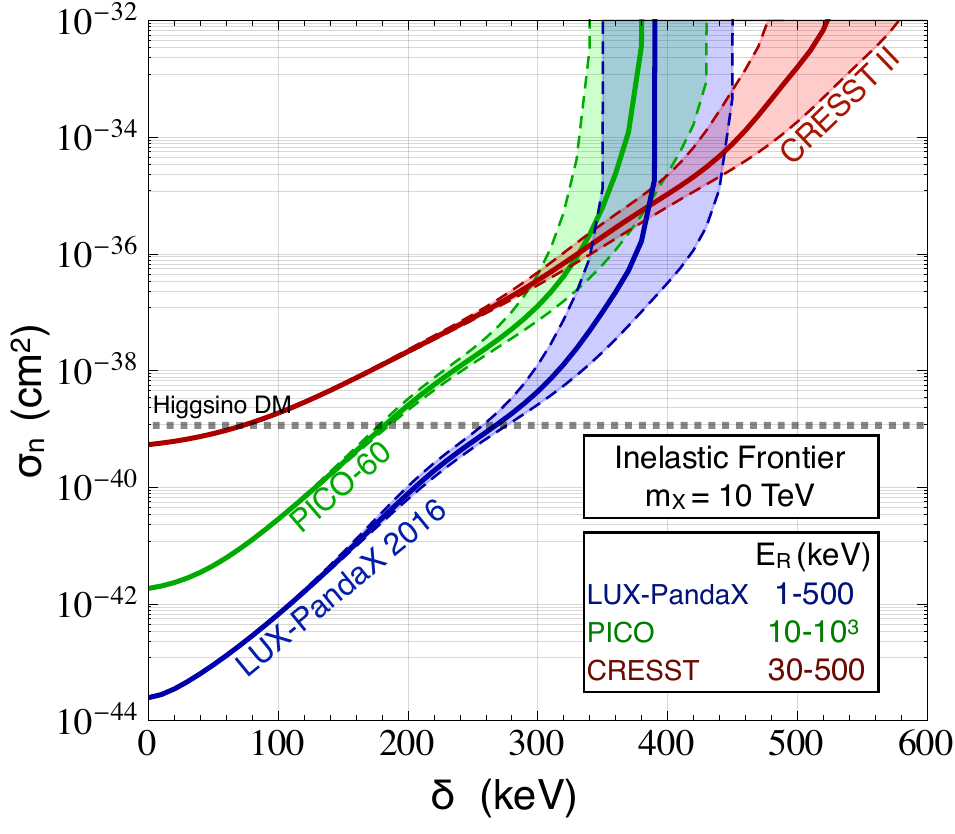}
\caption{Constraints on dark matter nucleon scattering ($90\%$ confidence), assuming integrated luminosities, event rates, and nuclear masses for LUX \cite{Akerib:2015rjg,LUXtalk}, PandaX II \cite{Tan:2016zwf}, PICO-60 \cite{Amole:2015pla}, and CRESST II \cite{Angloher:2015ewa}. Presently available recoil energy ranges  ($E_R$) used to derive bounds are indicated, along with extended ``inelastic frontier'' recoil energy ranges. The dotted horizontal line indicates the approximate Higgsino-nucleon inelastic cross-section for reference ($\sim 10^{-39} \; {\rm cm}^2$). The bands show how bounds vary within the 90\% confidence allowed values of the escape velocity given in \cite{Piffl:2013mla}, $v_{\rm esc} = 533^{+54}_{-41}~{\rm km/s}$.}
\label{fig:bounds}
\end{figure}

To find the \emph{currently} constrained $\sigma_n$ as a function of $\delta$, we integrate Eq.~(\ref{eq:rateER}) over the same energy region as the experiments, i.e. $1-30 \, \kev$ for xenon, $30 -120\, \kev$ for tungsten, etc.  For simplicity we assume the earth's speed was constant at 232 km/s during the whole period of data taking of each experiment.  In addition to the experimental parameters presented in Table~\ref{tab:experimentalnumbers} one must take into account that in the case of CRESST (PICO) only $\sim 64\%$ ($\sim 65\%$) of the target's mass is in the form of tungsten (iodine), and that each experiment (including LUX-PandaX) has a recoil-energy dependent efficiency to register nuclear recoils.  CRESST requires a simple rescaling of the exposure, whereas to account for the efficiency of the other experiments we rescale our bounds by an overall constant factor.  For PICO this rescale factor is chosen such that at $m_X=1$ TeV and $\delta=0~{\rm keV}$, our bounds agree with those presented by the PICO collaboration.  The ``LUX-PandaX 2016" curves are normalized to a $\sigma_{\rm n} = 2 \times 10^{-45} \; {\rm cm^2}$ bound at $m_X = 1 \; {\rm TeV}$ and $ \delta = 0 \; {\rm keV}$. For CRESST we use an overall efficiency of $64\%$ \cite{Angloher:2015ewa}.  As mentioned above, PICO varied its threshold energy throughout its period of data taking.  We approximate this behavior by using a fixed lower threshold of $E_R=10~{\rm keV}$, and again rescale to agree with the published PICO bounds at $\delta=0$ keV.  The resulting bounds on the scattering cross section, $\sigma_n$, are shown in the left two panels of Fig.~\ref{fig:bounds}. The bands correspond to varying the DM escape velocity within the $90\%$ confidence interval around its central value, $v_{esc} = 533^{+54}_{-41}\, \text{km/s}$. 

To determine the reach in the $(\sigma_n, \delta)$ plane that is obtainable by looking at high recoil energy data, we repeat the calculation but integrate up to $500\, \kev$ in recoil energy, using the same rescaling efficiencies as before. In the case of PICO, which collects events with recoil energies up to $\sim{1~\rm MeV}$, no improvement is possible.  For LUX-PandaX and CRESST, with no high-recoil background publicly available, we assume zero background events in the high energy bins, i.e that LUX-PandaX contains no events between $30-500\, \kev$, and CRESST II observes no events between $120-500\, \kev$ -- but the overall exposure and efficiency rescaling factors are kept the same. Since efficiencies are typically better at high recoil energy, where the bulk of signal events would reside for large $\delta$ dark matter, we anticipate that this rescaling will give conservative results. The resulting sensitivities are shown in the right hand panels of Fig.~\ref{fig:bounds} (including bands from varying $v_{esc}$), alongside the bound from PICO, which already employs high recoil data in its analysis.

Figure~\ref{fig:bounds} is one of the main result of this paper. Comparing the left and right hand panels and fixing $\sigma_n = 10^{-39}\, \text{cm}^2$ as a representative, weak-scale value, we see that current xenon constraints require $\delta \gtrsim 170\, \kev$; all current constraints (xenon and iodine) are satisfied once $\delta \gtrsim 220\, \kev$. Including higher-recoil data in the analysis would increase the sensitivity up to $300\, \kev$ (for weak scale $\sigma_n$), and similar jumps in the sensitivity occur for other $\sigma_n$ values. No single experiment dominates, despite the fact that the exposure in LUX-PandaX is a factor of $\sim 270$ times that of CRESST and 10 times that of PICO, further testament to the importance of heavy nuclear targets and high recoil data. We emphasize that the improved sensitivity, in the case of LUX-PandaX and CRESST, would come from evaluating data the experiments already have, and future data would of course improve their sensitivity.

\begin{figure}[t!]
\label{fig:modul}
\center
\includegraphics[width=0.48\textwidth]{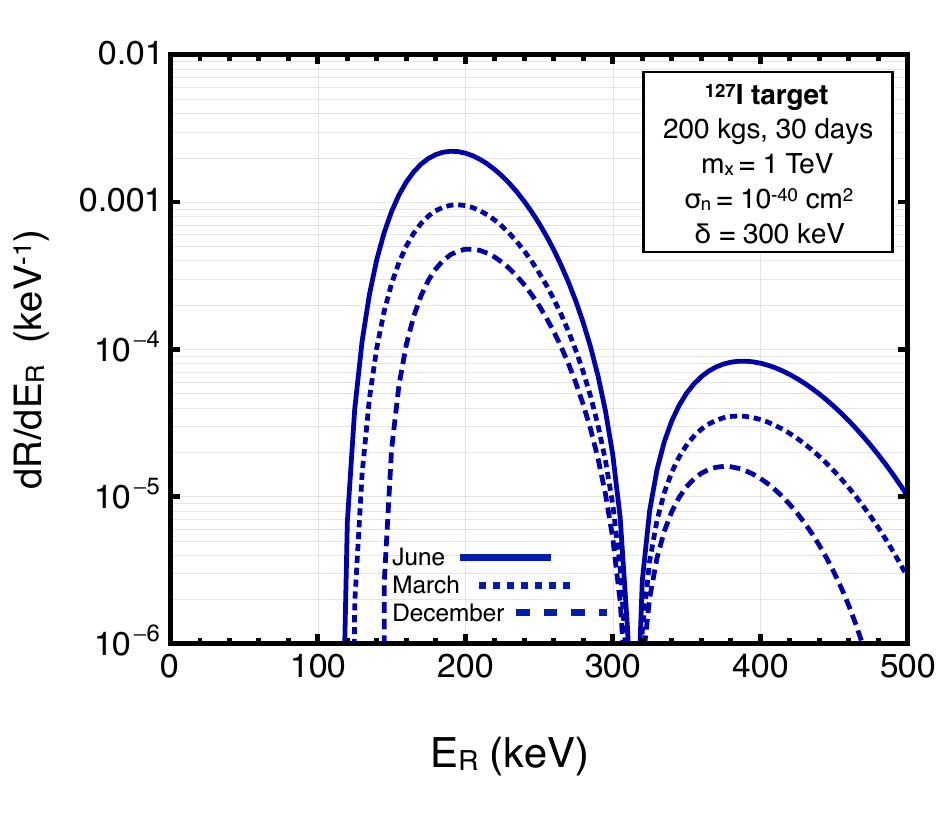}
\includegraphics[width=0.48\textwidth]{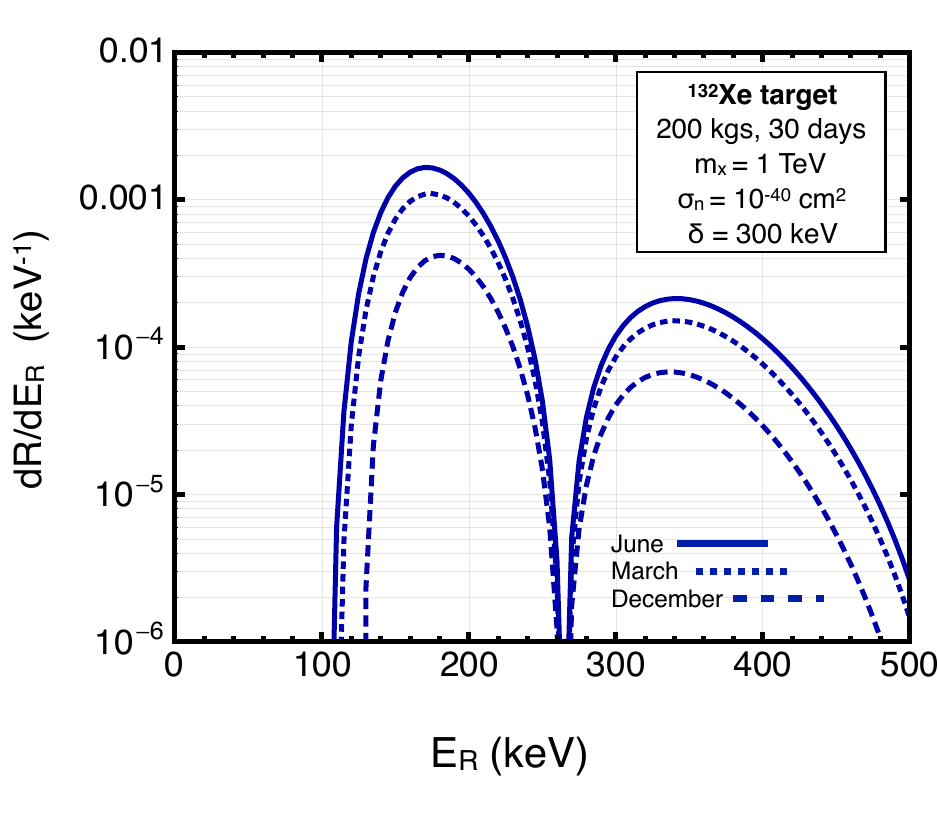}
\includegraphics[width=0.48\textwidth]{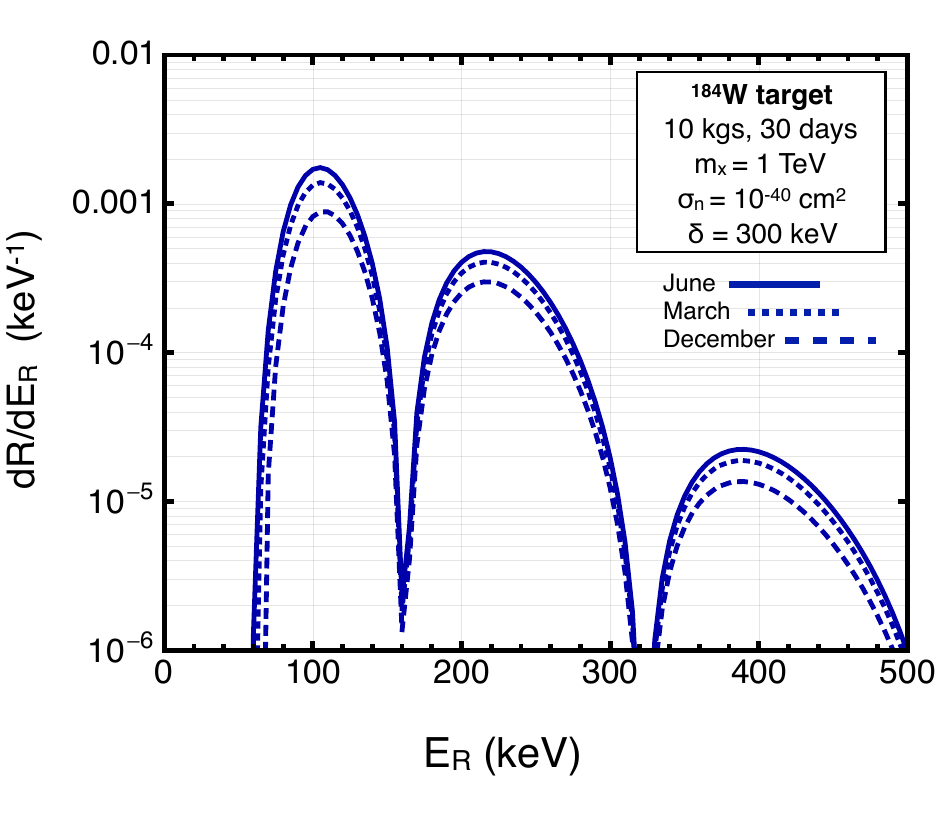}
\caption{Number of events per keV recoil energy for a 1 TeV mass WIMP ($\sigma_n=10^{-40}$ cm$^2$) with mass splitting of $\delta=300$~keV. As shown, the $\sim 300$~keV mass splitting results in a factor of $4$ yearly modulation in the number of dark matter scattering events in xenon and iodine. Note also that for a large inelastic mass splitting, a tungsten experiment with $1/20$ the target mass achieves comparable sensitivity to xenon and iodine experiments.}
\label{fig:mod}
\end{figure}

We also consider the impact that a yearly modulation of the average DM-Earth speed has on inelastic searches. The mass splitting reach and, in the case of dark matter detection, the spread of observed recoil energies, shift as the earth orbits the Sun, modulating the maximum lab frame WIMP-nucleon speed by $\sim \pm15~{\rm km/s}$ over the course of a year.  The higher the $\delta$, the greater the sensitivity to the tails of the DM distribution, and therefore the greater the sensitivity of the scenario to any modulation. An example of how the velocity modulation translates into a rate difference is shown below in Fig.~\ref{fig:mod}. There we fix $\delta = 300\, \kev$ for DM with mass $m_X  = 1\, \tev$ and per-nucleon cross section $\sigma_n = 10^{-40}\, \text{cm}^2$; to go from a velocity-averaged differential cross section into a rate we assume an exposure of 30 days and a target mass of 200 kg for iodine and xenon and 10 kg for tungsten. These target sizes were chosen to roughly coincide with conceivable future exposure. The large $\delta$ lies at the edge of the band of accessible LUX-PandaX parameter space in Fig.~\ref{fig:bounds}. Comparing the contours for December and June, we see there is a factor of $\sim4$ difference in the differential rate for iodine and xenon. For tungsten, the variation is smaller $\sim2$ because $300\, \kev$ is farther from the sensitivity limit on $\delta$. In addition to modulation in the rate at a given $E_R$, the range of permitted recoil energies also varies during the year, with a wider range being accepted around June 2nd.

Sizable modulation could be a useful tool to increase the significance of a small DM signal. Additionally, once DM is observed, modulation provides a way to obtain information on $\delta$.  For example, scattering occurring near the limit of xenon's inelastic sensitivity, will occur preferentially around June 2nd, and for incoming DM directed towards the Earth's velocity vector in the Galactic rest frame, $i.e.$ ${\rm cos} ~\theta_* \sim 1$, discussed in text surrounding Eq.~\eqref{eq:fullintegral}.  Thus, for a given $\delta$, future dark matter detectors with directional capability could improve searches for highly inelastic dark matter, by rejecting events around $\theta_* \sim \pi$ with too low a recoil energy, dependent on the time of year.  The details of how to extract $\delta$ out of an observed energy spectrum and modulation are left for future work.

\section{Models for high-recoil studies}
\label{sec:models}

So far, we have considered the implications of inelastic dark matter 
with a generic, model-independent cross section.  There are, however,
several constraints and implications that follow once a specific
model of the inelastic scattering cross section is considered.
In this section, we consider three well-motivated models that can contain
inelastic splittings of a size relevant for current and future
direct detection searches:  nearly pure Higgsinos; magnetic 
inelastic dark matter; dark photon mediated dark matter. 
In each of the models we consider, the two dark matter
states form a pseudo-Dirac pair, \ie\ they are Majorana fermions whose 
masses are slightly split,\footnote{There are also 
intriguing composite inelastic models involving excitations
from a (composite) scalar to a vector through dark photon 
mediation \cite{Alves:2009nf,Lisanti:2009am}.} 
whereas the mediator responsible for the 
DM-nucleus interaction is different.
We then demonstrate that it is possible to have large
inelastic scattering cross sections off nuclei, and negligible 
elastic scattering, while avoiding other pertinent constraints.  

In the Higgsino and dark photon models, the mass of the mediator ($Z$ and $Z'$ respectively) is set by the VEV of a scalar field that spontaneously breaks a gauge symmetry and is also responsible for the mass splitting between $X_2$ and $X_1$.  In the case of Higgsinos, the VEV is introduced by a doublet of Higgs bosons spontaneously breaking the Standard Model gauge groups SU(2)$_{\rm W} \times$U(1)$_{\rm Y} \rightarrow U(1)_{\rm EM}$, while in the case of dark photon mediated dark matter a scalar boson spontaneously breaks a dark U(1)$_{\rm D}$ gauge symmetry, which kinetically mixes with U(1)$_{\rm EM}$ of the Standard Model. In the magnetic inelastic dark matter (MIDM) model the mediator is the photon, and the model contains an $X_1-X_2$ transition magnetic dipole operator.

For each model, we are interested in the DM-nucleon cross section, $\sigma_n$ that renders the model free of any existing direct detection constraints, as well as the change in the allowed $\delta$ if the high-recoil searches advocated earlier are performed.  In addition to the allowed $(\sigma_n, \delta)$ parameter space, there are other constraints that must be checked before each model can be deemed viable.

First, as the relevant operator for inelastic scattering necessarily includes both $X_1$ and a heavier state $X_2$, one can insert this operator twice, $X_1 \to X_2 \to X_1$; once appropriately dressed with the SM particles, this allows for loop level elastic scattering. While these loop processes are suppressed by additional coupling and loop factors, they are free of the velocity suppression and recoil energy window mismatch present in the inelastic case. Therefore, it is important to estimate the size of elastic processes to determine the relevant parameter space where inelastic scattering does indeed dominate.

Second, our setup relies on the assumption that $X_1$ is the dominant component of DM, which clearly cannot be the case if $X_2$ lives too long. Even if all primordial $X_2$ has decayed to $X_1$, an $X_2$ component can be regenerated via DM collisions in our halo $X_1\, X_1 \to X_2\, X_2$. The number density of the regenerated $X_2$ is approximately \cite{Batell:2009vb}
\begin{equation}
n_{X_2} \; \sim \; n^2_{X_1}\, \tau_{X_2}\, \langle \sigma_{X_1 X_1 \to X_2X_2}\, v \rangle,
\label{eq:n2frac}
\end{equation} 
where $n_i$ is the number density of species $i$ and $\tau_{X_2}$ is either the lifetime of the excited state, or the lifetime of the universe (whichever is shorter).  The average kinetic energy of DM particles in the Galactic halo are large enough to allow this process to occur without suppression, potentially leading to a non-zero $X_2$ number density,
\begin{equation}
\frac{n_{X_2}}{n_{X_1}} \approx 4\times 10^{-12} \left(\frac{1\,\mathrm{TeV}}{M_{X_1}}\right)   \left(\frac{\tau_{X_2}}{\tau_U}\right)   \left(\frac{\langle \sigma_{X_1 X_1 \to X_2X_2}\, v \rangle}{3\times10^{-26}\mathrm{cm}^3\,\mathrm{s}^{-1}}\right) ~,
\label{eq:n2fraction}
\end{equation} 
here $\tau_U$ is the age of the universe.

 
 While the scattering of $X_1$ is endothermic and suppressed by kinematics, the scattering of $X_2$ is not, although the recoil spectrum is also somewhat peaked to high energies, so it is susceptible to the bounds on spin-independent \emph{elastic} scattering. The effective cross section for the regenerated $X_2$ population is $\sigma_{n,2} \sim \frac{n_{X_2}}{n_{X_1}}\, \sigma_{n,1}(\delta=0)$, where $\sigma_{n,1}$ is the per-nucleon (inelastic) scattering cross section for $X_1$.  Therefore, even a small fraction of sufficiently long lived or strongly interacting $X_2$ would be evident  in present-day direct detection experiments.\footnote{As an explicit example, if we assume 1 TeV DM and an inelastic per-nucleon cross section of $\sigma_{n,1} = 10^{-38}~{\rm cm^2}$, then $\frac{n_{X_2}}{n_{X_1}} $ must be $\leq 10^{-7}$ to avoid current bounds from LUX-PandaX.}

Finally,  we will focus on model parameters that yield the correct relic abundance whenever possible, but because the thermal history of the universe is not completely known (\eg\ there may have been late releases of entropy diluting over abundant DM) we will allow ourselves the liberty of considering other regions.

\subsection{Higgsinos}
\label{sec:higgsinos}
Higgsino dark matter can be concisely defined as a pair of fermions, which are doublets of SU(2)$_{\rm W}$ with hypercharge $\pm 1/2$. Higgsinos are typically studied as the fermion superpartners of two Higgs bosons in the minimal supersymmetric standard model (MSSM), see $e.g.$ \cite{Martin:1997ns,ArkaniHamed:2006mb}. The electrically-neutral component of the Higgsino doublets have the same quantum numbers as, and therefore mix with, other MSSM fermions, such as the singlet bino and SM weak triplet wino. If the wino or bino masses are much heavier than the Higgsino, as is the case in some ``split'' supersymmetric models \cite{Fox:2014moa,Nagata:2014wma}, the mass splitting between Higgsino states can be small enough that Higgsinos could be found in high recoil data already collected at direct detection experiments. 

After electroweak symmetry breaking, the Higgsino dark matter sector is composed of two neutral Majorana fermions ($X_1,X_2$) with inter-state mass splitting: 
\begin{equation}
\delta_{\rm \tilde H} \,\simeq\, m^2_Z\, \Big(\frac{\sin^2{\theta_W}}{M_1} + \frac{\cos^2{\theta_W}}{M_2} \Big) + \mathcal O\Big( \frac{1}{M^2_{1,2}} \Big) =  \left\{ \begin{array}{cc} 192\, \kev\, \Big(\frac{10^7\, \gev}{M_1}\Big) & M_2 \gg M_1 \gg \mu \\ 640\, \kev\, \Big(\frac{10^7\, \gev}{M_2}\Big) & M_1 \gg M_2 \gg \mu \end{array}\right.
\label{eq:higgsinosplitting}
\end{equation}
where $M_1$, $M_2$, and $\mu$ are the bino, wino, and Higgsino mass term, respectively and $\mu \sim m_{\rm X_1},m_{\rm X_2}$ in the parameter space of interest. Additionally, lest the reader think that narrow splittings for Higgsinos only occur in ``split'' supersymmetry models, from the form of Eq.~(\ref{eq:higgsinosplitting}) it is clear that it is possible to achieve a (fine-tuned) small splitting even if $\mu \sim M_1 \sim M_2$, provided $M_1$ and $M_2$ have opposite sign. The details of how the narrow splitting between the lightest two neutralinos arises will not concern the remainder of this discussion. 

\begin{itemize}
\item {\it Relic Abundance: } The relic abundance of neutralinos has been studied extensively. For simplicity, we will assume a spectrum where all superpartners other than the Higgsinos are decoupled to the point that the inter-Higgsino splitting shrinks to $O(100\, \kev)$. In this limit, the contribution to the energy density fraction of the universe from the Higgsinos is~\cite{ArkaniHamed:2006mb}
\be
\Omega h^2 = 0.10\, \Big( \frac{\mu}{1\, \tev} \Big)^2.
\ee
The correct abundance therefore requires Higgsinos masses of $\sim 1.1$ TeV, which we will use throughout this section.

\item {\it Cross section:} Higgsinos couple to nuclei via the $Z$ boson, and the dark matter-\emph{nucleus} cross-section in this case can be parameterized in terms of the Fermi coupling $G_F$ and the DM-nucleus reduced mass $\mu_N$, 
\begin{align}
\sigma_{\rm NX}^{\tilde{H}} \; = \; \frac{G_{F}^2 \mu_N^2}{8 \pi} \left(A-[2-4s^2_{\rm W}]Z \right)^2,
\end{align}
where, matching to Eq. \ref{eq:dsnde}, the effective per-nucleon cross-section is $\sigma_{\rm n} \sim 10^{-39} \; {\rm cm}^2$, with a precise value that depends upon the nucleus being scattered upon ($i.e.$ the ratio of ``$A$'' to ``$Z$''). In Fig.~\ref{fig:bounds}, the cross-sections for Higgsino scattering off nucleons in tungsten and xenon nuclei are indicated with a horizontal line. Thus, we see that Higgsinos with inelastic mass splittings up to $220$ ($300$)~keV could be excluded with presently available PICO data (analysis of LUX-PandaX high recoil data). Finally, a future tungsten-based experiment with much larger exposure than CRESST has the potential to probe Higgsino DM with mass splittings up to $\sim 550 ~{\rm keV}$. 

\item {\it Loop-level elastic scattering:} At the nucleon level, tree-level $Z$ exchange leads to Higgsino inelastic scattering with a cross section
\begin{align}
\sigma_{\rm n}^{\tilde{H}} \; \sim \; \frac{\pi\, m_n^2\, \alpha^2_W}{8\, m^4_W}\times (\text{velocity\, factor}) \sim 10^{-39}\; \text{cm}^2 \times (\text{velocity\, factor}), 
\end{align}
where $\alpha_W = g^2_W/(4\pi)$, $m_n$ is the nucleon mass, and we have assumed the DM mass is $\gg m_n$. In this case, the exchange of two gauge bosons leads to elastic scattering via a box diagram. As shown in Ref.~\cite{Hisano:2010fy, Hisano:2011cs,Hill:2013hoa,Hill:2014yka}, the double gauge boson exchange processes can be decomposed into twist-0 and twist-2 pieces, and these subprocesses are additionally suppressed by a factor of the nucleon mass (or momentum) divided by the $W^{\pm}$ mass (at amplitude level). Approximating the full loop-level cross section with the twist-2 $W^{\pm}$ exchange component, the result is
 \begin{equation}
 \sigma_{\rm n, loop}^{\tilde{H}}\; \sim \; \frac{m_n^4 \alpha^4_W}{\pi\, m^6_W}\, f_q^2 \sim 10^{-47} \; {\rm cm}^2,
 \end{equation}
where $f_q$ are hadronic matrix elements and are $\mathcal{O}(0.1)$. These factors render the loop-level elastic DM-nucleon scattering cross section closer to $10^{-47} {\rm cm}^2$. In further detail, as emphasized by Ref.~\cite{Hisano:2010fy, Hisano:2011cs,Hill:2013hoa,Hill:2014yka}, approximating the cross-section with the $W^{\pm}$ box diagram also overestimates the cross-section, as there is an additional, accidental cancellation between contributions from twist-0 and twist-2 operators. For $SU(2)_w$ doublet DM and $m_H = 125.7~{\rm GeV}$ these cancellations lead to $ \sigma_{\rm n, loop}^{\tilde{H}} \lesssim 10^{-48}\, \text{cm}^2$. 

\item {\it Exothermic bounds:} Comparing present direct detection limits to the tree-level inelastic Higgsino-nucleon scattering cross-section we see that the abundance of $X_2$ must satisfy
$n_{X_2}/n_{X_1}  \lesssim \; 10^{-6} \,{\rm TeV}/m_{X_1}$.  The Higgsino has a one-loop radiative decay $X_2 \rightarrow X_1 + \gamma$ with width (in the limit where all other superpartners are decoupled \cite{Haber:1988px})
$\Gamma_{X_2 \rightarrow X_1+\gamma}\, \sim\, \alpha_{em}\, \alpha^2_W\, \delta^3/(4\pi^2 m^2_X)$ and a weak scale scattering cross section,
$\sigma_{X_1X_1\rightarrow X_2X_2} \sim g_2^4/m_X^2$.
Thus, from Eq.~(\ref{eq:n2fraction}) it is clear that detectable Higgsino inelastic scattering will be endothermic.  
\end{itemize}

\subsection{Magnetic Inelastic Dark Matter}
\label{sec:MIDM}

We now consider a model which again has two Majorana fermions $\chi_1, \chi_2$ nearby in mass, $m_{\chi_2} = m_{\chi_1} + \delta$ but now their inelastic interaction with the SM is through a magnetic dipole operator~\cite{Chang:2010en, Kumar:2011iy,Weiner:2012cb, Weiner:2012gm},
\begin{equation}
\left( \frac {g_M}{4} \right) \frac{e}{2 m_{\chi}} \chi_2\, \sigma_{\mu\nu}\, \chi_1 F^{\mu\nu}.
\label{eq:magmoment}
\end{equation}
Since the DM is Majorana in nature there is no diagonal dipole operator, and only the transition dipole is allowed.  A perturbative UV completion \cite{Weiner:2012cb} of the theory generates this operator, with $g_M \sim m_\chi/(8\pi^2 M)$, after integrating out a heavy charged fermion and scalar of mass $M$ that couple to the DM.  However, we have chosen to adopt the operator normalization inspired by proton/neutron magnetic moments as would be expected if the DM was a composite of a new strongly coupled sector.  In such a model we expect $g_M\sim 1$. 

\begin{itemize}
\item {\it Relic abundance:} Above the scale of electroweak symmetry breaking the gauge invariant dipole operator involves the hypercharge field strength, $B_{\mu\nu}$.  Thus, in addition to the dipole operator with the photon (Eq.~(\ref{eq:magmoment})), one would expect a dipole operator involving the $Z$ boson, $-\tan\theta_W \left( \frac {g_M}{4} \right) \frac{e}{2 m_{\chi}} \chi_2\, \sigma_{\mu\nu}\, \chi_1 Z^{\mu\nu} $.  These interactions allow for DM annihilation into pairs of SM fermions, $W^+W^-$, and at higher order in the dipole coupling, annihilations to $\gamma\gamma$, $\gamma Z$, and $ZZ$.  The dominant annihilation is into up-type quarks and charged leptons, and for $m_\chi\sim 1\,\mathrm{TeV}$ and $g_M\sim \mathcal{O}(1)$ the annihilation cross section is sufficient for the DM to be a thermal relic.  For larger DM masses or smaller dipole moments the cross section is too small.  However, DM can still be a thermal relic if there are additional annihilation modes, for instance to a light dark photon which has a very small kinetic mixing with the SM photon.

\item {\it Cross Section:} The direct detection signal of MIDM is the inelastic collision of DM with the SM through exchange of a photon.  The DM couples to the charge and magnetic moment of the proton and the magnetic moment of the neutron.   These low energy couplings to nucleons take the form
\begin{equation}
\; \frac{1}{q^2}\left(\frac{e}{2m_n}\right)\left(\frac{e}{2m_\chi}\right) (\bar{\chi}_2\, i\, \sigma_{\mu\nu}\,q^{\nu}\, \chi_1)\left(\bar{p}\, P^{\mu}\, p + \frac{g_p}{2} \bar{p}\, i\, \sigma_{\mu\alpha}q^{\alpha}\, p +  \frac{g_n}{2} \bar{n}\, i\, \sigma_{\mu\alpha}q^{\alpha}\, n\right),
\end{equation}
where $g_p\approx 5.6\, (g_n\approx -3.8)$ are proportional to the magnetic moment of the proton (neutron) using the normalization of Eq.~(\ref{eq:magmoment}), and $P^{\mu} (q^\mu)$ is the sum (difference) of the momentum flowing through the nucleon line.  

To go from these low energy DM-nucleon couplings to the DM-nucleus differential cross section, we follow the formalism of~\cite{Fitzpatrick:2012ix,Anand:2013yka} augmented by the work of \cite{Barello:2014uda}, which extended the formalism to inelastic scattering. Specifically, having written the interaction as a sum over non-relativistic operators $\mathcal O_i$: $\mathcal L_{\chi-N} = \sum\limits_{\tau = p,n}\sum\limits_i c^{\tau}_i\, \mathcal O^{\tau}_i$ where the $\tau$ index allows different coefficients ($c_i$) for protons and neutrons\footnote{Or, equivalently, different coefficients for different nuclear isospin combinations}, we then insert these interactions inside a nucleus.  The final differential DM-nucleus cross section can be written as a sum over eight nuclear response functions, each weighted with a coefficient function that depends on the $c^{\tau}_i$ and kinematic factors such as the momentum exchange, the relative initial velocity, and the spin of the DM.  The nuclear responses are functions of the recoil energy alone; we obtain them from the Mathematica package associated with Ref.~\cite{Anand:2013yka}. Following Ref.~\cite{Barello:2014uda}, the sole consequence of the inelastic nature of the collision is a $\delta$ dependent shift in the coefficient functions.  For MIDM the dipole-charge interaction is due to non-zero coefficients of $\mathcal O_{1}$ and $\mathcal O_{5}$ in the notation of~\cite{Fitzpatrick:2012ix}, while the dipole-dipole interaction is through $\mathcal O_{4}$ and $\mathcal O_{6}$.

In Fig.~\ref{fig:gMplot} below we show how the direct detection bounds on MIDM change as a function of the $m_{\chi_2} - m_{\chi_1}$ mass splitting. As the formulation of $d\sigma/d E_R$ following Ref.~\cite{Anand:2013yka} has non-trivial velocity dependence, there is no simple analog of $\sigma_n$ (the per-nucleon cross section), therefore we can't easily combine the MIDM bounds with Fig.~\ref{fig:bounds}. Instead, we plot the bound on $g_M$ as a function of $\delta$ for several values of $m_{\chi}$.  At small $\delta$ the dominant nuclear responses are those associated with $\mathcal O_1, \mathcal{O}_5$ while at large $\delta$, 
$\mathcal{O}_{4},\mathcal{O}_{6}$ dominate.\footnote{We note that this an interesting example where  one of the non-trivial operators, $\mathcal O_5$ discussed in Ref.~\cite{Anand:2013yka} actually dominates over the `standard' operators $\mathcal O_1, \mathcal O_4$.}  This transition from spin-independent to spin-dependent and the associated change in nuclear responses explains why the present bounds from PICO are nearly comparable to frontier bounds from LUX-PandaX, despite their lower exposure.  However, at low $\delta$ the LUX-PandaX bound is considerably stronger than that coming from PICO.

\begin{figure}[t!]
\centering
\includegraphics[width=0.47\textwidth]{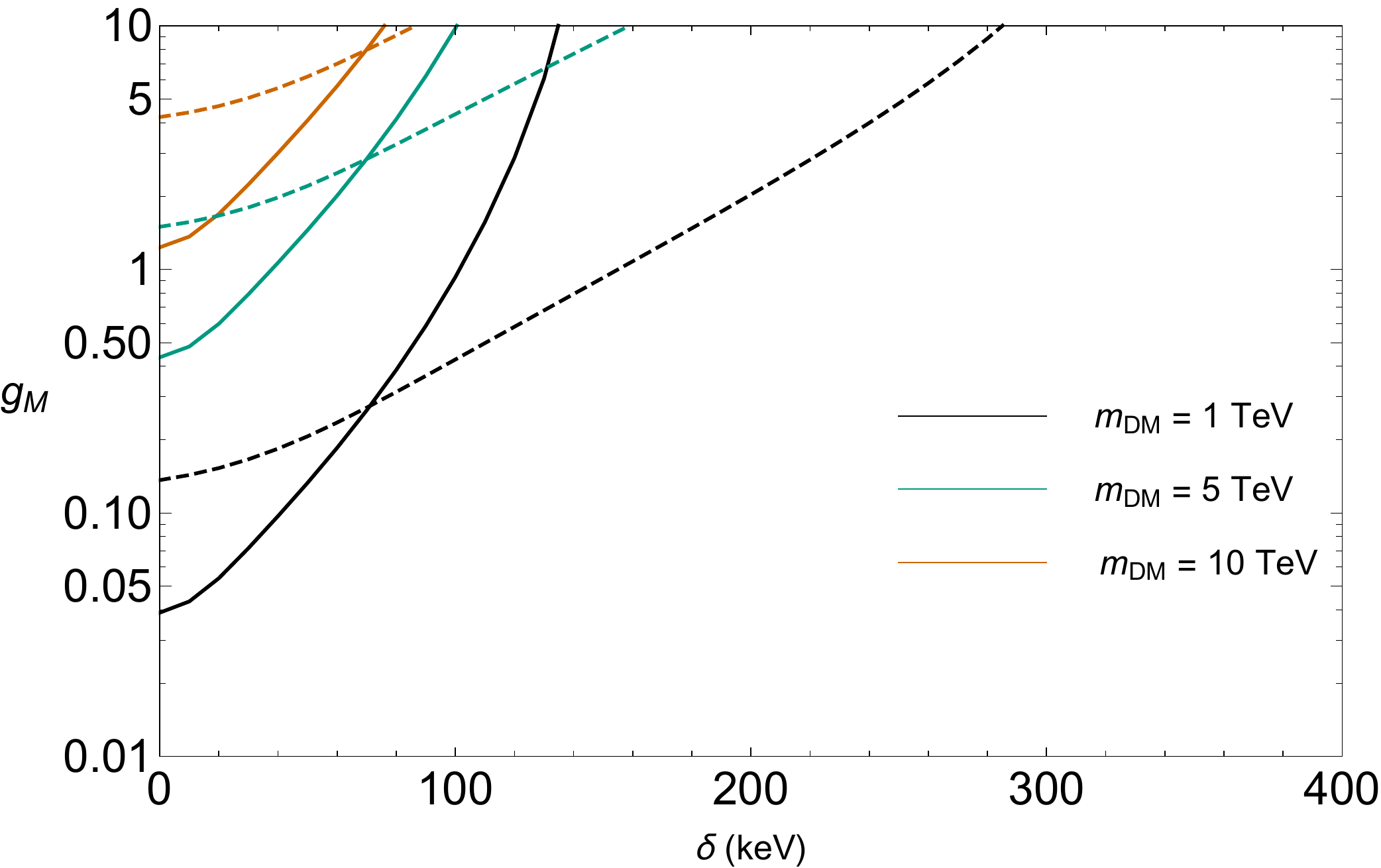}
\includegraphics[width=0.47\textwidth]{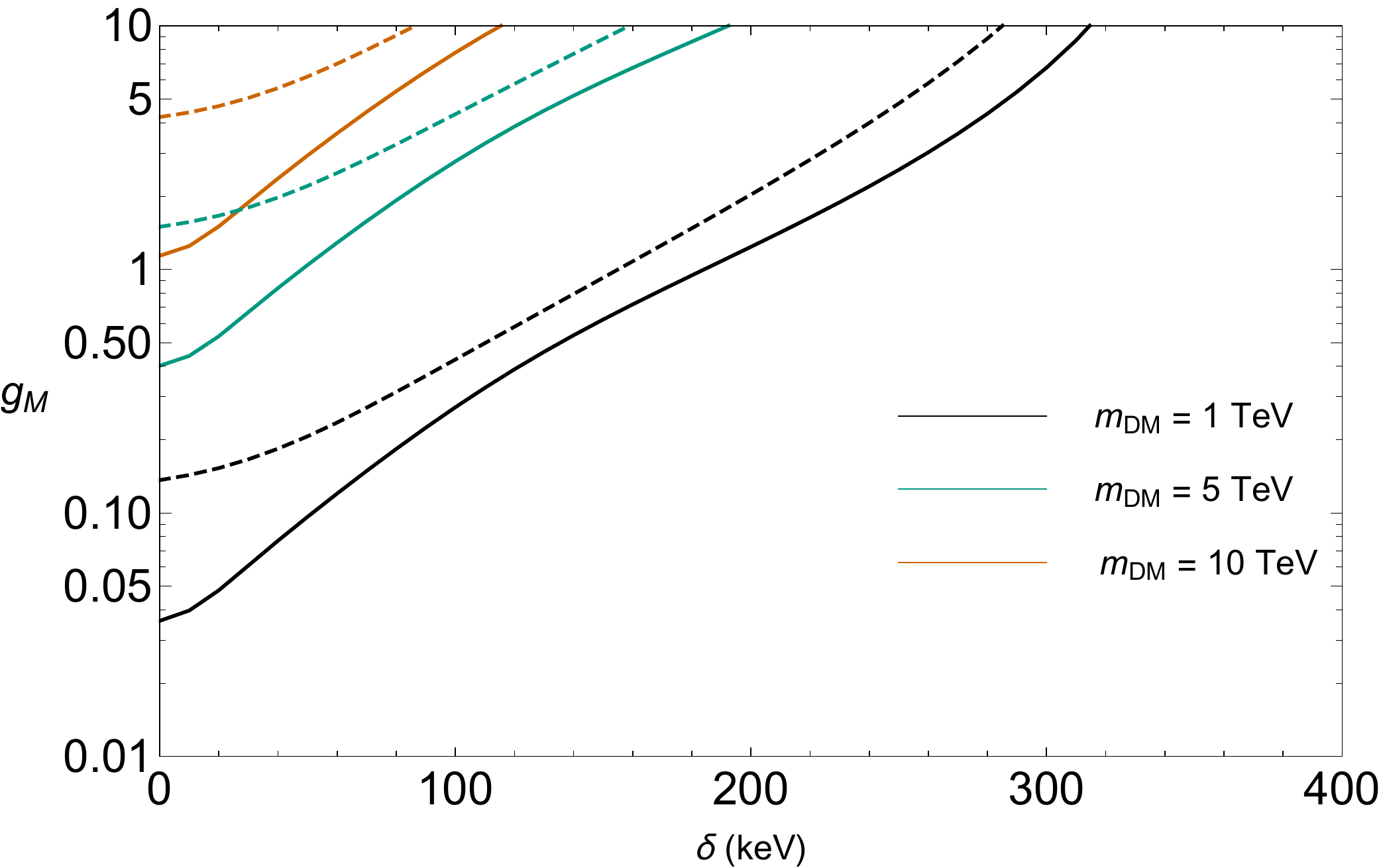}
\caption{Bounds on $g_M$ in the MIDM scenario as a function of mass splitting $\delta$, for DM masses of 1, 5, and 10 TeV. In the left panel we show the bounds from existing data, with dashed lines indicating the bounds from PICO and solid lines indicating the bounds from LUX-PandaX. In the right panel, we show the bounds if the ``inelastic frontier" is probed at LUX-PandaX using existing exposure. There is no competitive bound from CRESST as the exposure is too small, as is the abundance of the (stable) tungsen isotope with non-zero spin $^{183} W$.}
\label{fig:gMplot}
\end{figure}
\item {\it Loop-level elastic scattering: } Loop level elastic scattering can occur through two insertions of the MIDM operator.  This loop integral is log divergent and is cutoff at the scale which generates the DM dipole operator.  The effective operator generated by the loop process contains both an axial vector-axial vector interaction and a scalar-scalar interaction, the latter being suppressed by $m_n/m_\chi$.  Because of this suppression, for the DM masses we are interested in, the strongest bound comes from the spin-dependent operator and the per-nucleon scattering cross section is
\be
\sigma_{\rm n, loop}^{\rm MIDM} \sim \frac{\alpha_{em}^4}{\pi} \left(\frac{3\, g^2_M}{16\, m^2_\chi}\right)^2\, \mu_n^2~,
\ee
where $\mu_n$ is the DM-nucleon reduced mass.
Given the existing bounds on spin-dependent cross sections this constrains $g_M\ltap 500$ for 1 TeV DM, a very weak constraint.
\item  {\it Excited state lifetime:} The lifetime of the excited DM state is determined by the two body decay to $\chi \gamma$.  This width is $\Gamma(\chi_2 \to \chi_1\, \gamma) \sim \alpha\, g^2_M\, \delta^3/(2\, m^2_{\chi})$, so the excited state is very short lived and its abundance is tiny  (Eq.~(\ref{eq:n2fraction})).

\end{itemize}

\subsection{Dark Photon Mediated Dark Matter}
\label{sec:darkphotonsare_thebest}

Dark matter may couple to visible particles through a new massive vector boson, often referred to as a dark photon, that mixes with the hypercharge gauge boson \cite{Holdom:1985ag}. In these dark photon-mediated dark matter models \cite{Pospelov:2007mp,Cheung:2007ut,Feldman:2007wj,ArkaniHamed:2008qn,Finkbeiner:2009mi}, dark sector inter-state mass splittings arise if the scalar boson that spontaneously breaks the U(1)$_{D}$ gauge symmetry also couples to dark matter. The Lagrangian is given by
\begin{equation}
\mathcal{L} \; = \; \mathcal{L}_{\rm SM} + |D_{\mu} \Phi|^2 - V(\Phi) - \frac{1}{4} V_{\mu \nu}^2 + \epsilon V_\mu \partial_\nu F^{\mu \nu} + \bar{\psi} (iD_\mu \gamma_\mu - m_{\psi})\psi +(\lambda_{\rm D} \Phi\, \psi^T\, C^{-1}\, \psi + {\rm h.c.})
\label{eq:dplag}
\end{equation}
where $V,F$ are the $U(1)_{\rm D},U(1)_{\rm em}$ gauge bosons respectively, $D_\mu \equiv \partial_\mu + ie_{\rm D} V_\mu$, and $C$ is the charge conjugation matrix. Note that the dark matter particle $\psi$ is a Dirac fermion, with charge $e_{\rm D}$ under the $U(1)_{\rm D}$ gauge symmetry, which is half the $U(1)_D$ charge carried by $\Phi$. This permits the displayed Yukawa terms. Once $\Phi$ gets a VEV, the dark photon becomes massive, $\langle \Phi \rangle =v_{\Phi}$, and the Yukawa term induces a mass splitting between the two Majorana fermions composing $\psi$. The mass eigenstate Majorana fermions  ($X_1,X_2$) have mass $m_{X_{1,2}} = m_{\psi} \pm \delta_{X_i}$, where  
\begin{equation}
\delta_{X_i} \; \equiv \; m_{X_2}-m_{X_1} \simeq \lambda_{\rm D} v_{\rm \Phi} \; = \; 100 \; {\rm keV} \left( \frac{\lambda_{\rm D}}{10^{-3}} \right) \left( \frac{v_{\Phi}}{100~{\rm MeV}} \right).
\end{equation}

\begin{figure}[t!]
\center
\includegraphics[width=0.49\textwidth]{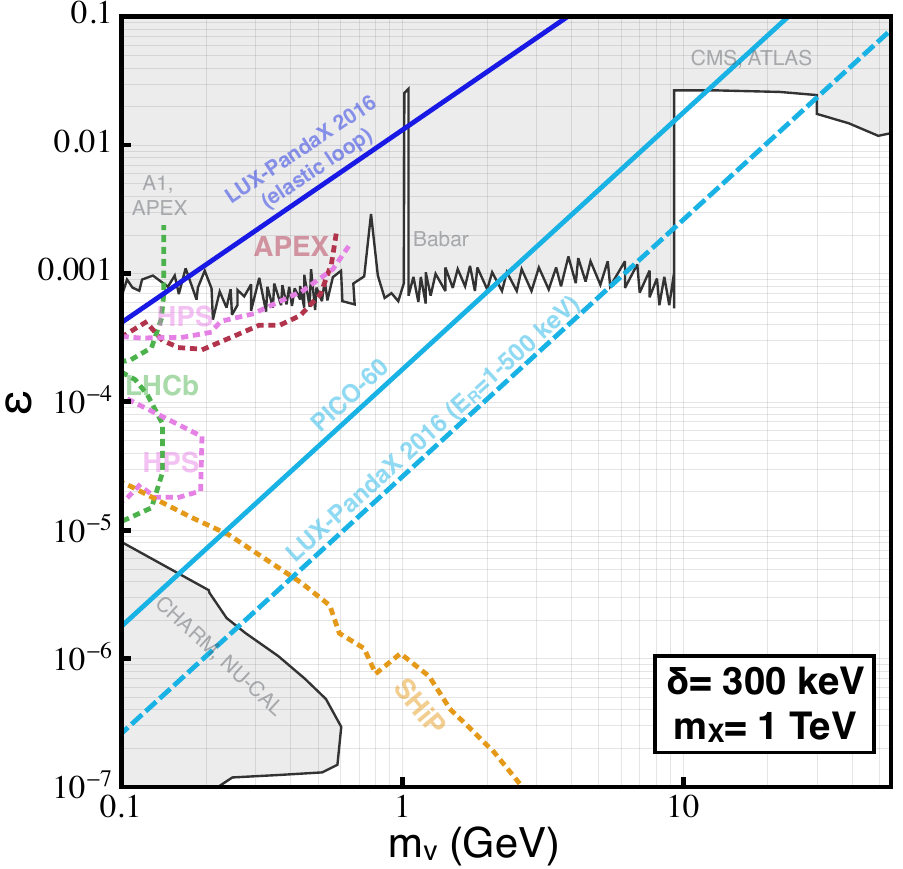}
\includegraphics[width=0.49\textwidth]{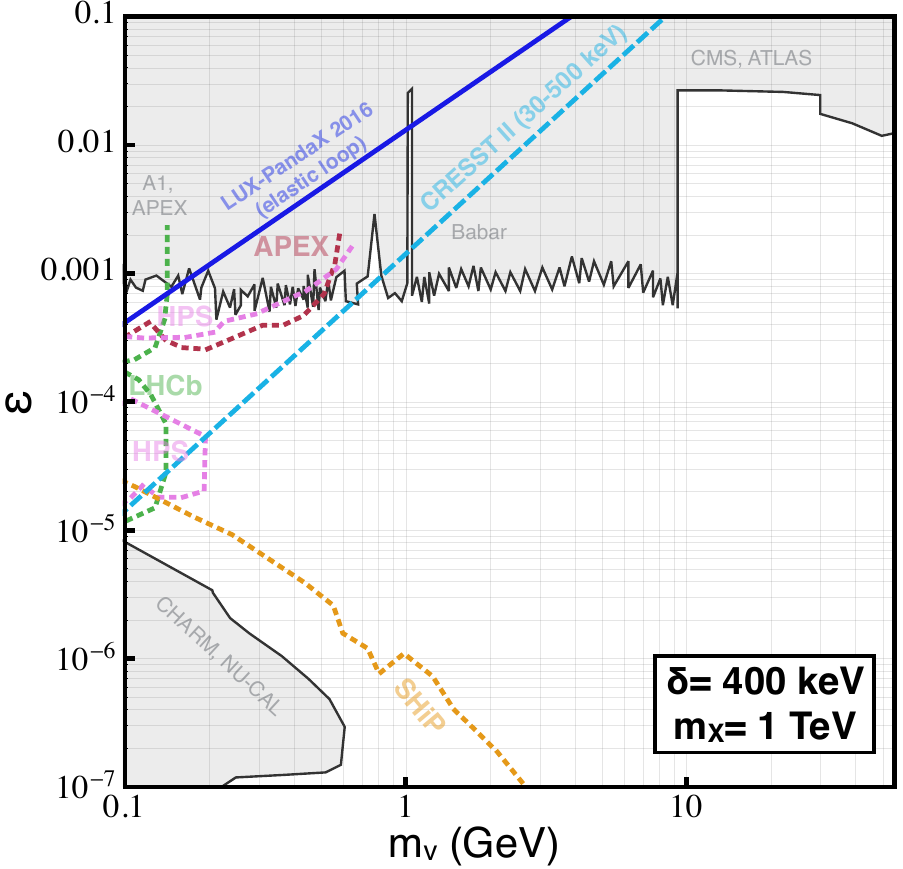}
\caption{Dark photon parameter space is shown, with inelastic bounds obtainable for 1 TeV mass, thermally-produced dark photon mediated dark matter, which fixes the dark sector gauge coupling $\alpha_{\rm D}$, as described in the text. The solid (dashed) cyan lines show the best present (future high recoil) bound on inelastic scattering of dark photon-mediated dark matter, for the inelastic mass splitting indicated in each figure. The dark blue line shows the bound LUX-PandaX \cite{Tan:2016zwf,LUXtalk} set on loop-induced elastic scattering of dark photon-mediated dark matter, as described around Eq.~(\ref{eq:loopdp}). Constraints on dark photons \cite{Abrahamyan:2011gv,Essig:2013lka,Goudzovski:2014rwa,Merkel:2014avp,Batell:2014mga,Curtin:2014cca} are shown in gray. Note that these gray constraint regions restrict dark photons in general, while the bounds derived here assume fermionic dark matter with a 1 TeV mass that freezes out to the observed relic abundance. Upcoming experimental searches for dark photons, complementing high recoil searches for dark photon-mediated inelastic dark matter ($e.g.$ \cite{Beranek:2013yqa,Alekhin:2015byh,Ilten:2015hya,Moreno:2013mja}) are indicated with dotted lines.}
\label{fig:dpbounds}
\end{figure}

\begin{itemize}
\item{\it Relic abundance:} The dark sector gauge coupling $\alpha_{\rm D}$ can be fixed by requiring that dark matter freeze-out to the observed relic abundance. In the limit that SM particles and the dark photon are much less massive than DM, $m_{V} \ll m_{X_{i}}$, the cross-section for non-relativistic DM-DM annihilation, $\left\langle\sigma_{\rm ann.} v\right\rangle \sim \pi \alpha_{\rm D}^2/2 m_{X_{1}}^2$.  Inserting this annihilation cross section into the standard WIMP annihilation freeze-out machinery \cite{Kolb:1990vq,Pospelov:2008jd} yields 
\begin{equation}
\alpha_{\rm D}^{\rm freeze-out} \; \simeq \; 3.7 \times 10^{-2} \left( \frac{m_{\rm X_1}}{{\rm TeV}} \right).
\label{eq:dpfo}
\end{equation}
We assume this value for $\alpha_{\rm D}$ throughout the remainder of this section.
\item{\it Cross section:} Dark Matter in this scenario will scatter inelastically off protons in nuclei by exchanging a dark photon, which mixes with the Standard Model photon and $Z$ boson after U(1)$_{\rm D}$  and electroweak symmetry are broken. For a nuclear target with atomic number $Z$, the DM-nucleus scattering cross-section is:
\begin{equation}
\sigma_{\rm N X}^{\rm D} \; = \; \frac{16 \pi \alpha_{\rm em} \alpha_{\rm D} \epsilon^2 \mu_N^2}{m_{\rm V}^4} Z^2,
\label{eq:dpcrosssectionN}
\end{equation}
where $\alpha_{\rm em}$ is the electromagnetic gauge coupling constant , $\mu$ is the reduced mass, and $m_{\rm V} \simeq  \sqrt{4 \pi \alpha_{\rm D}} v_{\Phi}$ is the mass of the dark photon.

Dark photons are under attack from a variety of experiments, whose constraints we summarize in Fig.~\ref{fig:dpbounds} in the $(m_V, \epsilon)$ plane.  In addition to existing bounds from the LHC, BaBar, etc. (shaded in Fig.~\ref{fig:dpbounds}) \cite{Abrahamyan:2011gv,Essig:2013lka,Goudzovski:2014rwa,Merkel:2014avp,Batell:2014mga,Curtin:2014cca}, we also indicate the anticipated bounds from upcoming experiments such as SHIP, LHCb, and APEX \cite{Beranek:2013yqa,Alekhin:2015byh,Ilten:2015hya,Moreno:2013mja}. To make this figure, we set the DM mass to 1 TeV, then use Eq.~(\ref{eq:dpfo}) to fix the $U(1)_{\rm D}$ coupling. 

Choosing two sample $\delta$ values, we can use the formalism of Sec. \ref{sec:rate} to calculate the sensitivities from current and potential high recoil searches and overlay them on the $(m_V, \epsilon)$ plane. In the left hand panel, $\delta = 300\, \kev$, and we see the upper left triangle of parameter space, roughly from $\epsilon > 3\times 10^{-7}$ for  $m_V = 0.1\, \gev$ to $\epsilon > 0.1$ for $m_V = 100\, \gev$ could be ruled out by high recoil xenon studies. Analysis of current data estimated in this study, weakens this bound by roughly an order of magnitude, with the strongest limits coming from PICO (dashed, cyan line). Increasing $\delta$ to $400\, \kev$, shown in the right panel of Fig.~\ref{fig:dpbounds}, no currently existing study can provide a bound. However, future analysis of high recoil tungsten data can exclude $\epsilon \ge 10^{-5}$ for $m_V = 0.1\, \gev$ to $\epsilon \ge 0.1$ for $m_V = 10\, \gev$. For reasons explained in the next bullet point, we do not consider dark photon masses lighter than $0.1$ GeV in Fig.~\ref{fig:dpbounds}.

\item{\it Loop-level elastic scattering:} Generally, the loop-induced elastic cross-section depends on the mass of the dark photon. For dark photons heavier than the momentum exchange, $m_V \gtrsim 100~{\rm MeV}$, the effect is comparable to the Higgsino case but with $m_V$ replacing $m_W$ and $\alpha_W \rightarrow \sqrt{\alpha_D\alpha_{\rm EM} \epsilon^2}$. This leads to the following expression for the nucleon level, loop-induced elastic scattering of dark photon-mediated dark matter,
\begin{equation}
\sigma_{\rm n,loop}^{\rm D} \; \sim \; \frac{\alpha_{\rm D}^2 \alpha_{\rm em}^2 \epsilon^4 m_n^4 f_q^2}{\pi m_V^6},
\label{eq:loopdp}
\end{equation}
where we take $f_q \sim 0.1$ as for Higgsinos. The bound on loop-induced elastic scattering of dark photon-mediated dark matter is shown in Fig.~\ref{fig:dpbounds} for parameter space where $m_V \gtrsim 100~{\rm MeV}$. For $m_V \lesssim 100~{\rm MeV}$, dark photon exchange becomes long-range compared to the nucleus and the cross section is enhanced by a factor of $Z^2$ compared to Eq.~\eqref{eq:loopdp}; see Ref.~\cite{Batell:2009vb} and \cite{Weiner:2012cb} for further discussion. As an enhanced loop-level elastic scattering rate dramatically reduces the parameter space where inelastic scattering is relevant, we will focus on $m_V \gtrsim 100\, \mev$

\item {\it Exothermic bounds:} For the dark photon-mediated model specified around Eq.~\eqref{eq:dplag} and for $\delta_{X_i} \sim 100$~keV mass splittings, the largest contribution to $X_2$'s decay width is from $X_2 \rightarrow X_1 + 3 \gamma$ \cite{Batell:2009vb}. In the parameter space of interest, this leads to decay times in excess of $\tau_{X_2 \rightarrow X_1 + 3 \gamma} \gtrsim 10^8$~yrs and $n_{X_2}/n_{X_1} \sim 10^{-4}$. For TeV-mass dark photon-mediated dark matter, which for $\epsilon \sim 10^{-4}$ and $m_V \sim 0.1$~GeV has a per-nucleon cross-section $\sigma_{\rm n} \sim 10^{-32} \; {\rm cm}^2$, this relative abundance is too large for predominant endothermic scattering (given LUX-PandaX constraints on $\sigma_{\rm n}$) by about eight orders of magnitude. Therefore, the inelastic frontier studied here only applies to the subset of dark photon scenarios where additional decays permit $X_2 \rightarrow X_1 + \text(something)$ to happen within a year or less. One possibility is if the scalar responsible for breaking the dark $U(1)$ is light enough to permit the two-body decay $X_2 \rightarrow X_1 + \Phi$. However, one must then determine whether $\Phi$ decays promptly enough to avoid spoiling big bang nucleosynthesis and other cosmological complications. A second possibility is to extend the dark photon model to include an inelastic magnetic dipole operator, as discussed in Section \ref{sec:MIDM}, which would imply a magnetic inelastic cutoff $\Lambda/e \equiv 8 m_X/(g_M\, e) \lesssim 10^6 {\rm ~GeV} (\delta/{100 \; \rm keV})^3$. Figure \ref{fig:dpbounds} displays bounds on dark photon mediated dark matter assuming that $n_{X_2} \ll n_{X_1}$.
\end{itemize}

\section{A Hint of Inelastic Dark Matter at CRESST}
\label{sec:cresstwimp}

In CRESST's most recent publication of their results
\cite{Angloher:2015ewa}, they observed four events
in their data well below the $5\sigma$ lower boundary
of the expected background from electron/gamma leakage.
Two (four) of these events are within $1.5\sigma$ ($3\sigma$)
of the anticipated tungsten nuclear recoil light yield.
The CRESST collaboration suggests these are an
additional source of un-vetoed $\alpha$ background.

\begin{figure}[t!]
\center
\includegraphics[scale=1]{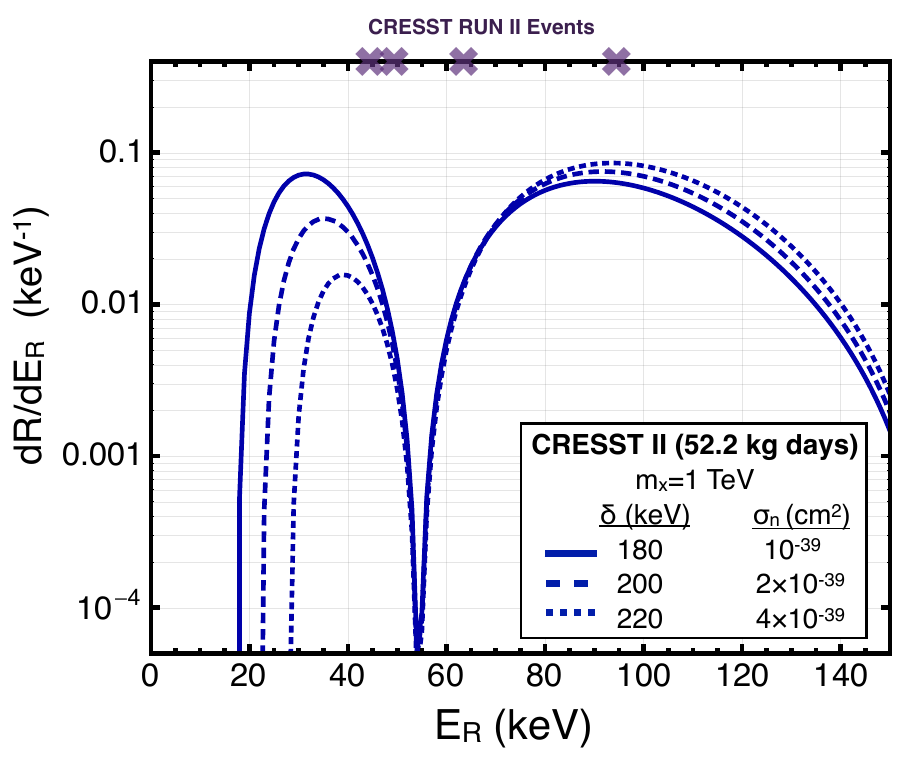}
\caption{The expected scattering rate for 1 TeV mass WIMP dark matter with a $\delta = 180,200,220$~keV inter-state mass splitting, and nucleon scattering cross-sections indicated, which were determined by requiring 4 events, assuming a tungsten target and 52 kg days of CRESST data, with the Earth-Milky Way relative velocity fixed to March 1st. The expected rate curves are compared to the 4 high recoil events recently observed at CRESST \cite{Angloher:2015ewa}.}
\label{fig:cresstwimp}
\end{figure}
 
While these events may be due to nuclear backgrounds, it is amusing to consider instead that these four events
arise from inelastic dark matter with a large mass
splitting. In Fig.~\ref{fig:cresstwimp}, four high nuclear recoil events recently observed at the CRESST experiment are compared to the expected rate from a 1 TeV mass WIMP dark matter particle, with inter-state mass splittings of $180,200,220$~keV, and nucleon scattering cross-sections fixed to produce four events at CRESST. First, it is interesting to note that the two events at CRESST which are closest to the ``form-factor zero'' at $E_R=55$~keV, are also those outside the $1.5\sigma$ tungsten nuclear recoil light yield band (compare Fig.~\ref{fig:cresstwimp} to Fig.~5 in \cite{Angloher:2015ewa}). Therefore, these two events may be disfavored both as genuine tungsten scattering events and as dark matter (because the rate severely declines around $E_R=55$~keV). Of course this assumes that the Helm form-factor we employ accurately predicts where the expected rate drops to zero.  More precise nuclear shell model calculations for tungsten (such as done by Ref.~\cite{Klos:2013rwa,Vietze:2014vsa} for xenon and germanium) would further clarify a detailed interpretation of these events as dark matter scattering.

Next we note that for the inelastic mass splittings shown in Fig.~\ref{fig:cresstwimp}, our high recoil analysis of PICO-60's results displayed in Fig.~\ref{fig:bounds}, would be in tension with a dark matter explanation of the 4 events observed at CRESST. Nevertheless, Fig.~\ref{fig:cresstwimp} demonstrates how the expected rate changes as a function of recoil energy, and how this can be used to discriminate between dark matter signal events and background. Moreover, this tension motivates a re-examination of whether PICO-60's cut on low acoustic power events, used to reject the  $\alpha$ decay background, might have removed signal events, a possibility discussed in the PICO-60 document \cite{Amole:2015pla}.

\section{Conclusions} 
\label{sec:conclusions}

In this paper we have analyzed inelastic dark matter across the full range of mass splittings that are kinematically accessible in terrestrial fixed target scattering experiments.  One of our main results, Fig.~\ref{fig:bounds}, shows that the strongest constraints on the parameter space ($\delta$, $\sigma_n$) currently arise from xenon at LUX-PandaX, iodine at PICO, and tungsten at CRESST.  CRESST has the best bounds at the largest presently probed inelastic splittings, $\delta \simeq 300$-$375$~keV, due to its use of tungsten.  A modest increase in their recoil energy range would allow CRESST to extend their sensitivity up to $\delta \simeq 550$~keV\@.  PICO's unique experimental design and analysis does not impose an upper bound on the nuclear recoil energies, and so PICO can place the best constraints on intermediate inelastic splittings, $\delta \simeq 160$-$300$~keV\@.  Their experimental setup (when they use iodine) is optimal for discovering inelastic dark matter in this intermediate regime.  LUX, PandaX and Xenon100 have upper bounds on their recoil energy (and do not even show data above about $30$-$50$~keV), and this limits their sensitivity to smaller inelastic splittings $\delta \lesssim 160$~keV\@. 

We have also shown that xenon experiments can obtain much better sensitivity if they analyze their data up to much higher recoil energies.  For example, with their existing exposure, LUX-PandaX could be sensitive to scattering cross sections $30-40$ times larger than PICO by analyzing nuclear recoils up to $500$~keV\@.  This may require calibrating S2/S1 discrimination for electron versus nuclear light yields at energies higher than presently studied.  However, since Xenon10 has calibrated up to $300$~keVnr (using an AmBe source \cite{Angle:2009xb,Sorensen:2011bd}), it appears entirely possible to do so.

Increasing the sensitivity to the inelastic frontier will probe theories with truly ``weak'' interaction cross sections.\footnote{In effect, inelastic dark matter can put the ``W'' back into ``WIMP''!}  One of the amazing properties of inelastic dark matter is that it is possible to have cross sections that are fundamentally weak scale $\sigma_n \sim 10^{-39} \; {\rm cm}^2$ or larger, but existing experiments are not sensitive due to both the very small local WIMP velocity distribution that is probed, combined with the relatively small recoil energy range considered by many experiments (LUX, Xenon100, CDMS).  Indeed, the ``race for the bottom" (the focus to obtain the highest sensitivity at the lowest recoil energies) has certainly increased sensitivity to light, elastically-scattering dark matter, but provides no benefit to inelastic dark matter.

We have also discussed several specific models of dark matter that predominantly scatter through an inelastic interaction. Three examples are Higgsinos, magnetic inelastic dark matter, and dark photon mediated dark matter, which all predict small splittings between the dark matter and its neutral excited state.  The common element of all of these models is a Dirac fermion that is split into two Majorana states.  The lighter Majorana fermion is the dark matter, with highly suppressed elastic interactions with nuclei.  Instead, the $\chi_1 \rightarrow \chi_2$ transition has a much larger cross section, and for some models ($e.g.$, nearly pure Higgsinos) may be the only way to detect the dark matter through direct detection experiments.  Composite models can also provide a scattering cross section that is dominated by an inelastic transition.  This was explored in the context of a dark photon transition in Ref.~\cite{Alves:2009nf,Lisanti:2009am}.

A novel aspect of some models of inelastic dark matter
is the possibility that the dominant nuclear response could
\emph{change} as $\delta$ is increased.  Magnetic inelastic dark matter
depends on several nuclear responses, with the contributions from
spin-independent and spin-dependent being the dominant ones.
At larger inelastic splittings, the spin-dependent contribution
becomes the dominant contribution to the scattering rate.  
As a consequence, all other things considered equal, the experiment 
that is most sensitive to high inelastic splittings is the one 
that has elements with the largest spin-dependent coupling.  
This is the reason why the existing constraints from PICO
are very nearly comparable to the \emph{projections} from LUX-PandaX
(if they were to extend their analysis to high recoil energies, 
as we have already emphasized).  Namely, iodine has a much
larger spin-dependent coupling than xenon, giving PICO an
intrinsically better sensitivity despite their analyzed 
exposure being much lower than LUX-PandaX. 

Another unusual aspect of inelastic dark matter is the 
possibility of a large annual variation in the direct
detection rates.  It is well known (see $e.g.$
\cite{TuckerSmith:2001hy,TuckerSmith:2004jv,Chang:2008gd})
that annual modulation is significantly enhanced
in inelastic dark matter models.  This is because 
the kinematically accessible part of the velocity distribution 
is significantly reduced, and thus more sensitive to the Earth's motion, 
than elastic dark matter.  Very large inelastic dark matter
can have very striking annual variation of the direct
detection rates.  Since the kinematics are sculpted by the
inelastic threshold, as discussed in Sec.~\ref{sec:sbasics},
both the lower and upper bound on the recoil energy, 
reached by going to the highest velocities, have large 
annual variation of scattering rates.  This characteristic 
variation is completely distinct from the annual modulation 
of vanilla elastic scattering, and provides an 
excellent probe to uncover the inelasticity 
(i.e., the mass difference $\delta$) 
associated with dark matter.  

Finally, we have seen that nuclear responses at higher recoil
energies are a crucial component to extracting sensitivities
for inelastic dark matter.   As emphasized in 
Secs.~\ref{sec:rate}, \ref{sec:MIDM}, and \ref{sec:cresstwimp}, 
at the high recoil energies relevant for large inelastic mass splittings,
there are characteristic recoil energies at which the rate for 
dark matter scattering off nuclei drops to zero, as a consequence 
of destructive interference between the effective size of the nucleus 
and the momentum exchanged between the dark matter and nucleus. 
These ``form factor zeroes'' may be utilized to discriminate a genuine 
dark matter signal from background. However, this also shows how 
important accurate nuclear scattering form factors are for the 
inelastic frontier. Modern nuclear shell model calculations 
have characterized the nuclear response of spin-independent and 
spin-dependent scattering off xenon and germanium
\cite{Klos:2013rwa,Vietze:2014vsa}; 
it would be advantageous to have a similar level of precision
applied to iodine and tungsten.
This would allow better estimates of the theoretical errors on 
high recoil nuclear response functions. 

There is a rich opportunity for discovering dark matter on the 
inelastic frontier.  Continued and improved studies by direct detection
experiments are poised to verify (or significantly constrain)
inelastic dark matter through analyses of high recoil data combined 
with increased exposures of heavy elements.

\section*{Acknowledgements}

We thank 
P.~Agrawal,
Y.~Bai,
B.~Batell,
S.~Chang,
E.~Dahl,
A.~Delgado,
O.~Harris,
K.~Howe,
C.~Newby,
T.~Roy  
and
P.~Sorensen  
for useful discussions. We are also extremely grateful to A.~Fitzpatrick and W.~Haxton for clarifications of Ref.~\cite{Anand:2013yka} and for providing us with the latest set of response functions.  JB thanks Los Alamos National Laboratory (LANL) and the Center for Theoretical Underground Physics (CETUP) for hospitality while this work was completed.  PJF, GDK, and AM are grateful to the Mainz Institute for Theoretical Physics (MITP) for its hospitality and its partial support during the completion of this work.  PJF thanks the Alexander von Humboldt Foundation for support during the completion of this work. This work of GDK was supported in part by the U.S. Department of Energy under Grant No. DE-SC0011640. The work of AM was partially supported by the National Science Foundation under Grant No. PHY-1417118.  Fermilab is operated by Fermi Research Alliance, LLC under Contract No. DE-AC02-07CH11359 with the United States Department of Energy.  

\appendix

\section{Form factor details}
\label{app:formfactor}

In this study we employ a variety of different form factors. When considering the model-independent bounds on inelastic spin-independent scattering in Sec.~\ref{sec:prosp}, we use the Helm form factor for iodine and tungsten targets:
\begin{equation}
F_{\rm h}^2(E_{\rm R}) \; = \; \left(\frac{3 j_1(q r)}{q r} \right)^2 e^{-s^2q^2},
\label{eq:helm}
\end{equation}
where $q = \sqrt{2 E_{\rm r} m_{\rm N}}$, $r=\sqrt{r_n^2 -5s^2}$, $s=1~{\rm fm}$, and $r_n=1.2 A^{1/3} ~{\rm fm}$. For scattering off xenon, we employ form factors obtained from shell-model simulations of xenon isotopes (see e.g. Ref.~\cite{Fitzpatrick:2012ix}),
\begin{equation}
F_{\rm v}^2(E_{\rm R}) \; = \; \frac{e^{-u}}{A^2} \left(A + \sum_{n=1}^5 c_n u^n \right)^2,
\label{eq:hax}
\end{equation}
where $u=q^2b^2/2$, $b^2=m_n^{-1}(45A^{-1/3}-25A^{-2/3})^{-1}~{\rm MeV^{-1}}$, and the coefficients $c_n$ for xenon isotopes are given in Ref.~\cite{Vietze:2014vsa}. For example, in the case of $^{132}$Xe, $c_1= -132.841$, $c_2= 38.4859$, $c_3= -4.08455$, $c_4= 0.153298$, $c_5= -0.0013897$.

Form factors come up again when we study magnetic inelastic DM (MIDM) in Sec.~\ref{sec:MIDM}. There, the presence of additional velocity dependence in $d\sigma/dE_R$ means there is no analog of a per-nucleon cross section and we need to derive bounds using scattering off the whole nucleus. To properly include the nuclear response in this case, we use the response functions provided in the Mathematica notebook accompanying Ref.~\cite{Anand:2013yka}. For the interactions present in MIDM, only a subset of the 8 possible response functions (specifically, $\mathcal W_M, \mathcal W_{\Sigma'}, \mathcal W_{\Delta}$ and $\mathcal W_{\Delta\Sigma'}$) are required. See Ref.~\cite{Anand:2013yka} for details of the $\mathcal W_i$ and Ref.~\cite{Barello:2014uda} for modifications required when considering inelastic scattering.

\bibliographystyle{JHEP.bst}

\bibliography{highmassinelastic}

\end{document}